\documentclass{aa}  

\usepackage{graphicx}
\usepackage{txfonts}
\usepackage{lipsum}
\usepackage{subcaption}        
                                
\usepackage{lscape}             
\usepackage{ulem}                
\usepackage{placeins}            
\usepackage{pdfpages}                                

\newcommand*{\rom}[1]{\expandafter\@slowromancap\romannumeral #1@}
\newcommand\kmsec{km~s$^{-1}$}
\newcommand\masyr {mas~yr$^{-1}$}
\newcommand\dexkpc {dex~kpc$^{-1}$}
\newcommand\teff{$T_{\fontsize{6}{6}\selectfont \mbox{eff}}$}\normalfont

\begin{document}

   \title{The internal kinematics and chemistry of 20 Milky Way strings}

   \author{
        C. Boeche\inst{1}
        \and
        A. Vallenari\inst{1}
        \and
        J. Alonso-Santiago\inst{2}
        \and
        A. Bragaglia\inst{3}
        \and 
        R. Carrera\inst{3}
        \and
        E. Dalessandro\inst{3}
        \and
        E. Carretta\inst{3}
        \and
        E. Franciosini\inst{2}
        \and
        A. Frasca\inst{2}
        \and
        S. Lucatello\inst{1}
            \and 
        D. Romano\inst{3}
        \and
        G. Andreuzzi\inst{5,6}
        }
   \institute{INAF - Osservatorio Astronomico di Padova, vicolo dell’Osservatorio 5, 35122 Padova, Italy
   \and
    INAF – Osservatorio Astrofisico di Catania, via S. Sofia 78, 95123 Catania, Italy
   \and
   INAF - Osservatorio di Astrofisica e Scienza dello Spazio di Bologna, Via P. Gobetti 93/3, 40129 Bologna, Italy
   \and
   INAF – Osservatorio Astrofisico di Arcetri, Largo E. Fermi, 5,
   50125 Firenze, Italy
   \and
    INAF - Osservatorio Astronomico di Roma, via Frascati 33, 00178 Monte Porzio Catone, Italy
    \and
    Fundación Galileo Galilei – INAF, Ramble José Ana Fernandez Pérez 7, 38712 Breña Baja, TF, Spain}

   \date{Received }

  \abstract
   {The recent discovery of filamentary stellar structures in the Milky Way
disk raises the question of their formation in the context of
the Galaxy evolution.}
   {In this work, we aim to kinematically and chemically characterize several filamentary stellar structures by looking
for clues about their origin and formation mechanism.} 
   {Recent works using machine-learning techniques and $Gaia$ data have
identified many previously unknown Galactic strings.  We cross-matched such
data with the kinematics derived from the most recent $Gaia$ DR3 catalog as well as chemical data provided by the Galactic Archaeology with HERMES (GALAH) survey.}
   {We found that most strings contain one or more open clusters and
are kinematically hotter than open clusters (with internal velocity
dispersions in the 1--7~\kmsec\ range) but cooler than field stars.  18 objects
appear chemically homogeneous ($\sigma{[Fe/H]}\leq 0.1$~dex), while
two strings have a broad [Fe/H] distribution similar to 
the field stars but exhibit a different mean metallicity, which may identify 
them as migrating stellar structures. For seven objects the GALAH observations 
focus on the embedded open clusters; therefore, the chemical abundances are not 
representative of the whole structure. An object shows clues of chemical gradients 
along its extension in Galactic longitude.}
   {These objects may be evaporating or disrupted open clusters in which
the chemical homogeneity is preserved in an unbound and expanding cloud of
stars or co-moving stars belonging to a former and now disrupting star formation hub.
In a few cases they appear to be moving groups with no chemical homogeneity,
likely formed by the dynamical action of the Galactic bar or spiral arms.
More data are needed to confirm such hypotheses.}


   \maketitle
\nolinenumbers

\section{Introduction}

It is known that the Galactic disk hosts substructures that can be coherent
in space and velocity, clumping in the 6D phase space.
The search for and identification of such substructures has recently been boosted
thanks to the excellent astrometry of the $Gaia$ satellite
and the growing use of machine learning techniques. Many recent works successfully
coupled $Gaia$ DR2/DR3 data \citep{gaiaDR2,gaiaDR3} and machine-learning algorithms
and identified well-known open clusters as well as thousands of previously unknown open clusters (OCs; gravitationally bound) and associations of stars (gravitationally unbound with a common motion)
\citep[e.g.,][]{sim,cantat-gaudin2020A, castro-ginard, hunt-reffert, perren}. In these cases, we refer to such objects as co-moving stellar structures.
In other cases, groups of stars are unrelated, co-moving (i.e., clump in the velocity
space), unbound, and may be scattered in the physical space. These
are known as moving groups (MGs), as defined by \citet{eggen}. While OCs and 
associations are known to be the result of star
formation episodes, MGs can form dynamically \citep{dehnen, antoja}. 
In addition to OCs, associations, and MGs, new elongated filamentary stellar
ensembles were recently discovered. Some of them appear to be bridging neighboring
OCs \citep{jerabkova2019, beccari2020}. 
Some seem to embed many different OCs in one
large structure \citep{wang}. Others stand alone, covering
hundreds of parsecs in length and just a small fraction of it in thickness, such as the
Meingast~1 structure discovered by 
\citet[][hereafter M19]{meingast}; see also \citet{curtis} and \citet{hawkins}.
All these structures have age estimates of a few tens of
megayears (Myrs) and are believed to be coeval.

Because gas filaments are also found in giant molecular clouds 
\citep{myers}, hundreds of parsecs long
\citep{ragan}, it is straightforward to hypothesize 
that these strings are the relics of star formation episodes in filaments \citep{jerabkova2019}.
On the other hand, the disruption of an OC such as the Hyades with tidal tails hundreds of parsecs long 
\citep{meingast_hyades,roser} forms an elongated structure of stars similar to the
stellar filaments recently found.

\citet[][hereafter KC19]{kounkel}  identified 1900 structures in
the astrometric, photometric, and kinematic data of the $Gaia$ DR2
catalog. They were looking for high-density stellar regions with
the Python implementation of the hierarchical density-based spatial
clustering of the application with noise HDBSCAN \citep{hdbscan} in the
5D phase space: Galactic coordinates $l$ and $b$, parallax $\pi,$ and proper
motions $\mu_l$ and $\mu_b$ converted into kilometers per
seconds (\kmsec). They selected stars
located at low Galactic latitude ($|b|<30$ deg) and within one kiloparsec (kpc; $\pi > 1$~mas)
from the Sun. These structures are labeled as ``strings'' and ``groups'' by KC19 in their Table~2. The former are so named because they have filamentary, string-like shapes (width-over-length ratio from 1:6 to 1:80 according to KC19), while the latter are roundish and
compact\footnote{Throughout the paper we use the generic word ``object" to refer to both strings and groups; the word ``string" is used only where appropriate.}. Many of the groups are already known OCs, associations, or co-moving groups of stars,  
while 328 elongated structures were classified as strings. 
KC19 claimed that strings are roughly parallel to the Galactic plane and appear to be coeval
with a distribution in the color--magnitude diagram that can be fit with
a single isochrone. Besides, they do not seem to be disrupted OCs as
they do not show a higher density central region and because many of them
have ages much shorter than the tidal-disruption timescale.

Recently, 12 structures were identified by
\citet[][hereafter R25]{ratzenboeck1} in the solar neighborhood with low internal velocity
dispersions. Four of them are among the structures identified by KC19, while eight are new.
\citet{manea} studied 18 objects from the KC19 catalog, ten of which are strings.  They tested the chemical homogeneity of these objects with the
precise chemical abundances of the DR3 catalog of the GALAH Survey
\citep[Galactic Archaeology with HERMES,][]{galahDR3},
and concluded that all but one show a homogeneity that is higher than the
field stars and comparable to that of the open cluster M~67. Given the chemical homogeneity observed, they suggested 
that such structures may be coeval.
However, some caution is needed. \citet{zucker} showed that by moving from
$Gaia$ DR2 to more precise $Gaia$ DR3 astrometry \citep{gaiaDR3}, the dispersion of the stars
around the string's spine does not diminish as it would if they were
real objects. Given their unbound nature, Zucker and colleagues argued that
their apparent age is greater than their dispersing time, and therefore such
structures should not exist. Besides this, they showed that a 
chemically homogeneous group of stars can be fabricated by randomly drawing
stars from different open clusters, proving that the coeval origin of these
objects cannot be claimed. 

The origin of these structures
is not clear, their properties are often unknown, and even their existence is questioned. However, since they are located in the proximity of the Local and Sagittarius-Carina Arms, they are relevant to the study of disk star formation history and the dynamical processes in the solar vicinity.
Here we studied the internal kinematics and the chemical abundances of 20 objects with the aim of verifying whether these kinematically detected structures are also chemically related. This can provide constraints on their nature. Our sample includes 17 objects from KC19, two from R25, 
plus the Meingast~1 from \citet{ratzenboeck} data (hereafter R20).
We investigated their chemical homogeneity
using the precise chemical abundances (from spectra with resolution R$\sim28000$) of the GALAH DR4 catalog \citep{galahDR4} and the relation
between chemistry and internal kinematics using the astrometric data $Gaia$ DR3.

\section{Data}

Because the stars of the KC19 catalog (also called the Theia catalog)
and the Meingast~1 object
refer to the $Gaia$ DR2 catalog, we used the table 
{\it gaiadr3.dr2\_neighborhood} (provided by the $Gaia$ archive) to find the
best cross-match to the more recent DR3 $Gaia$ catalog \citep{gaiaDR3}. 
We collected the astrometry (position, parallax, and proper motions $\mu_{\alpha}$ and
$\mu_{\delta}$),  
radial velocity (RV), and photometry of the stars from the $Gaia$ DR3 catalog.
There was no need to do it for the R25 objects because they were identified from the $Gaia$ DR3 catalog.
We used the \citet{BJ21} catalog in order to have robust distance
estimates useful for data exploration. 
We computed Galactic longitude, $l$;
latitude, $b$; and proper motions in \masyr\ and in \kmsec\
(following what was done in KC19), as well as Galactocentric positions (X, Y, Z).
This computation was performed using the Astropy package \citep{astropy}.
For some stars \citet{BJ21} did not provide any distance, and therefore no
(X, Y, Z) position or velocity in \kmsec\ could be computed.

For the chemical abundances, we used the GALAH DR4 catalog. 
For GALAH stars, we followed the quality-selection recommendations
given on the GALAH DR4 website\footnote{https://www.galah-survey.org/dr4/flags/}.
We only chose stars with the quality flags {\it flag\_sp} $= 0$, {\it
flag\_sp\_fit} $= 0$, {\it flag\_red} $= 0$, {\it snr\_px\_ccd3 $>$ 30}, and {\it
flag\_fe\_h} $= 0;$
and chemical abundances with the quality flag {\it flag\_x\_fe} $= 0$ (where
$x\_fe$ represents the general abundance [X/Fe]). Furthermore, we only selected stars
with \teff\ between 4500\,K and 6500\,K to avoid possible systematics that affect the chemical abundances in too cold or too hot stellar atmospheres.
The GALAH survey provides elemental abundances with respect to iron
([X/Fe]) for up to 30 elements plus iron abundance [Fe/H]. 
However, the abundances of these elements have different precisions;
therefore, we decided to only consider elements that have a mean 
error in abundance $\overline{\sigma}_{[X/Fe]} < 0.08$~dex ([Fe/H] included). 
These are Na, Mg, Si, Ca, Sc, Ti, Cr, Mn, Fe, Co, and Ni.
For our analysis we decided to use elemental abundances
with respect to Hydrogen ([X/H]), which were derived as\\
$$
[X/H] = [X/Fe] + [Fe/H],
$$
while the errors are computed as
$$
\sigma_{[X/H]} = \sqrt{\sigma_{[X/Fe]}^{2} + \sigma_{[Fe/H]}^{2}}.
$$
We also used the GALAH RV {\it RV\_galah\_1} whenever there was no Gaia
RV available.

\subsection{Sample selection}\label{sec_samples}
By cross-matching the GALAH stars with the different catalogs,
we found 1269 stars in common with the Theia catalog: 42 stars for the R25
catalog, and 14 stars for Meingast~1. Keeping only the objects with at least
ten GALAH observed stars for a fair statistic, we were left with 18
Theia objects, Meingast~1,  and the two R25 objects S6 (whose core corresponds to Theia~371) and S7 (whose core corresponds to Theia~118), 
for a total of 20 distinct objects. Since Theia~118 is also present among the
18 Theia objects, in order to avoid
redundant data we dropped this Theia sample and used the S7 sample.
In addition, in the KC19 catalog, Theia~371 does not reach the minimum of ten GALAH stars of the
above conditions, while for the R25 catalog it does. See Table~\ref{tab:objects}
for an overview. Out of a total of 20 objects, only 17 are classified as string objects by KC19, whose authors reported an estimation of their length; the other three are classified as roundish objects, and no length was estimated.
\begin{table*}[!th]
\caption{Properties of the objects studied in this work. All the values are derived from KC19 with some exceptions explained in the notes.}
\label{tab:objects}
\centering
\setlength{\tabcolsep}{1mm} \begin{tabular}{llrrrrrrlr}
\hline\hline 
Name & other & is  &Log &  length & distance\tablefootmark{b} & reference & Number of\\
     & name\tablefootmark{a}  & string & Age   &  (pc) & (pc) &  & OC/MG\tablefootmark{c}\\
\hline
Theia 13 & Orion & y & 6.89 & 176.5 & 387.2 & KC19 & 13\\
Theia 613 & NGC~2516 & y & 8.4 & 349.6 & 404.4 & KC19 & 2\\
Theia 74 & NGC~2547 & y & 7.5 & 154.1 & 339.7 & KC19 & 7\\
Theia 1531 & Ruprecht~147 & n & 9.4 &  & 313.0 & KC19 & 1\\
Theia 143 & Trumpler~10 & y & 7.75 & 206.9 & 423.5 & KC19 & 6\\
Theia 369 & Pleiades & n & 8.13 &  & 143.1 &KC19 & 3\\
Theia 1004 & Hyades & n & 8.8 &  & 49.0 & KC19 & 1\\
Theia 216 & Theia~216 & y & 7.99 & 87.8 & 228.1 & KC19 & 2\\
Theia 823 & Mamajek~4 & y & 8.62 & 161.3 & 443.3 & KC19 & 1\\
Theia 598 & Alessi~9 & y & 8.46 & 160.8 & 221.0 & KC19 & 2\\
Theia 55 & NGC~2232 & y & 7.39 & 317.5 & 290.8 & KC19 & 9\\
Theia 452 & Theia~452 & y & 8.23 & 486.2 & 439.6 & KC19 & 4\\
Theia 97 & Theia~97 & y & 7.52 & 193.0 & 401.3 & KC19 & 5\\
Theia 1415 & Theia~1415 & y & 9.23 & 350.5 & 651.8 & KC19 & 0\\
Theia 309 & Theia~309 & y & 8.04 & 192.8 & 271.9 & KC19 & 0\\
Theia 301 & AB~Dor & y & 8.03 & 192.6 & 117.6 & KC19 & 5\\
Theia 908 & Theia~908 & y & 8.79 & 135.0 & 185.5 & KC19 & 0\\

S7 & NGC 2451A & y & 7.69 & 330 & 185 & R25 & 2\\
S6 & OCSN 87 & y & 8.20 & 300 & 140 & R25 & 2\\

Meingast 1 &  & y & 8.08 & 400 & 109 & R20 & 12\\
\hline
\end{tabular}
\tablefoot{
\tablefoottext{a}{This column refers to the region or OC associated with the objects according to KC19.}
\tablefoottext{b}{For the Theia objects the distances are derived from the heliocentric positions X, Y, and Z provided by KC19. For S6, S7, and Meingast~1 the distance is derived as the  mean position in X, Y, and Z of the stars members.}
\tablefoottext{c}{The number of the OC or MG embedded in each object is derived by cross-matching the objects' members to the HR24 catalog (see Sect.~\ref{sec_methods}).}
}

\end{table*}
For each object we wanted to study its star members, but also the surrounding
field stars  (i.e., stars that are located near the object that do not
belong to it) for comparison purposes.
To do so, we defined two samples for each object: the object sample and the test (that is, field-star) sample.
For kinematic studies the object sample is composed of all its members that
have $Gaia$ DR3 data, while for the chemistry studies we used a subsample with GALAH chemical data.
The test sample of an object was selected
by considering all the GALAH stars
found in the smallest volume (in the space, $l$, $b$, and parallax, $\pi$) that
frames all the stars belonging to the object\footnote{For Theia~1004 we 
had to enlarge such a  box by one extra degree in $b$ and $l$ and expand the
parallax interval to cover the distance between 10 and 140~pc in order to
have enough stars for the analysis.}. 
The sample so selected holds 
stars that belong to both the object and the nearby field. After
rejecting the stars belonging to the object, we took the stars left as the test sample. Finally, we used the constraint $[\alpha/Fe]<0.2$~dex in order to select only thin-disk stars.

\section{Methods}\label{sec_methods}

According to KC19, some of the objects are associated with known
OCs. We repeated the search for associated OCs using the updated OCs catalog
by \citet{hunt-reffert2024} (hereafter HR24). 
To find the OCs associated with the objects, we cross-matched the object
members with the OC members that have membership probability of $prob>0.5$ computed by HR24. Thus, the associated OCs are the ones for which the
matching star number is at least 30\% of the total number of members of
the OC considered\footnote{This condition was given in order to have a
reasonable number of OC members.}. This allowed us to analyze
the stars belonging to the OCs and the ones that belong to the host objects separately and compare them to each other.
In Figs.~\ref{diagnostic1}, \ref{diagnostic2}, and \ref{diagnostic3} we show 
the spatial shape and the velocity distributions of the objects for the $Gaia$ and GALAH
samples, with overplotted positions of the known associated OCs found.

\subsection{Chemistry}
For each object, we considered the chemical
abundances distribution and compared it to its test sample
to evaluate the chemical homogeneity and similarity with respect to the local field stars. 
Detailed information on the means and standard deviations of the chemical abundance distributions for each object and its test sample is reported in Table~\ref{tab_abd}. 
This is shown graphically in
Fig.~\ref{fig_XH_distr_mean_std_diff}. In some cases, the stars observed by
GALAH all belong to the hosted OCs. In such cases, the chemical analysis
characterizes the chemistry of the OCs to be distinguished from the chemistry of the object.
We also used the t-test (to compare the 
abundance means) and the Levene's test (to compare the spread of the distribution as standard deviations,
$\sigma$s) between the object and its test sample. These tests were
implemented in the Python {\it scipy.stats} packages
\citep{statsmodels}. 
They provide probability values (p-values) for the null hypothesis
(i.e., the two samples were drawn from the same population). When the p value was smaller than 0.05 we rejected the null hypothesis; i.e., the two samples belong to different populations. 

\begin{figure}[t]
\centering
\includegraphics[width=9cm, clip]{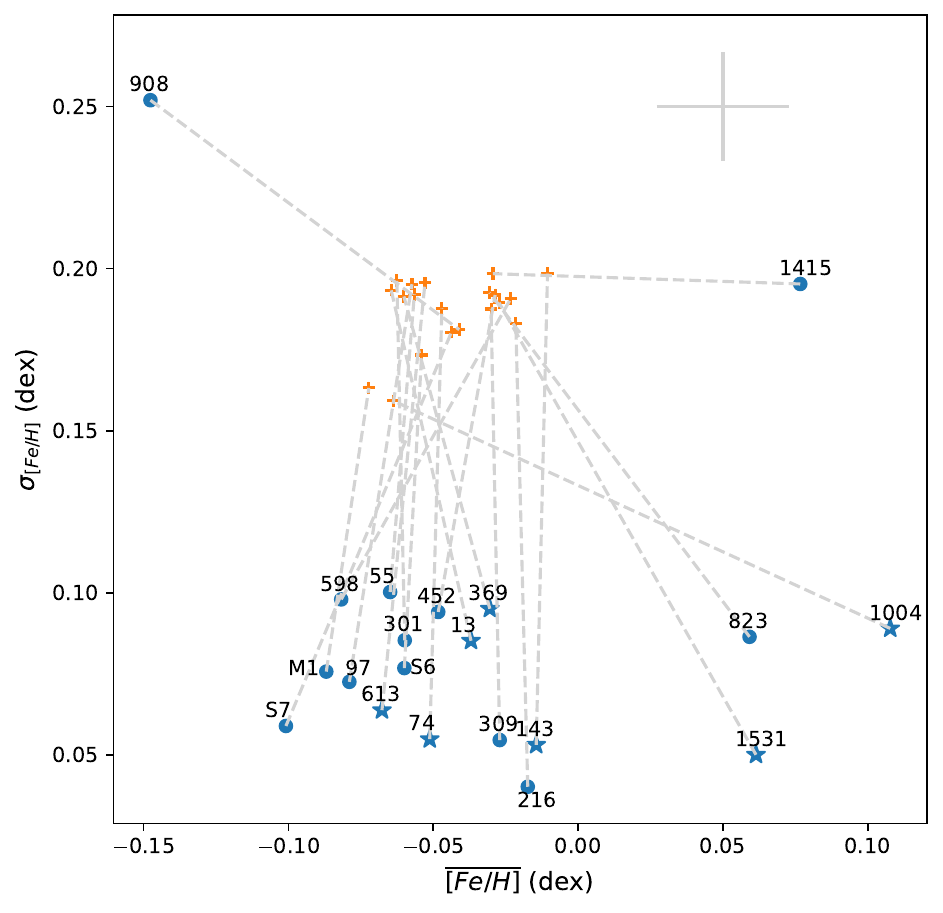}
\caption{Mean [Fe/H] (x-axis) and its standard deviation (y-axis) for the objects
(blue symbols) and their test samples (orange plus symbols) connected with a dashed gray
line. The blue star symbols represent the objects whose chemical abundances
describe only the associated OCs (and not the object as a whole). 
The blue circles represent the objects whose chemical abundances describe the whole structure. 
In the top right corner we report the mean errors for the x and y axes. 
The numbers label the structures.} 
\label{fig_mean_std}
\end{figure}

The means and standard deviations of [Fe/H] are
summarized in Fig.~\ref{fig_mean_std}, where we compare such values
for objects and their test samples. One can see that the objects and
their test samples are located in different areas of the plane: most of the objects
have a metallicity similar to their test samples (except for a few cases, which are discussed
later) but have a narrower iron-abundance distribution. 

\subsection{Kinematics}\label{sec_kinematic}
We used the kinematic information held by the $Gaia$ DR3 catalog 
to study the internal velocity dispersions of the OCs and their host
objects. The proper motions $\mu_{\alpha}$ and $\mu_{\delta}$ were corrected
for the virtual expansion due to the relative velocity between the objects
and the Sun using the formulas provided by \citet{brown1997}. The required RV of the object was computed as the mean Gaia RVs of the available
object members. The proper motions were converted into tangential velocities ($v_t$ hereafter) expressed in \kmsec. We define the internal velocity dispersion as the median absolute deviation (MAD) of the tangential velocities
as the mean of the two MADs $v_t^{\alpha}$ and $v_t{\delta}$:
$$
MAD_{v_t} = \frac{MAD_{v_t^{\alpha}} + MAD_{v_t{\delta}}}{2}.
$$  
Similarly, we used the $MAD_{RV}$ for the dispersion along the line of sight.
MADs were used to express the internal velocity dispersion of an object ($MAD^{obj}$), of its test sample ($MAD^{test}$), or of an
OC/MG ($MAD^{OC}$)\footnote{Among the structures identified by HR24 that we found, part of the objects there are OCs and MGs. For the sake of brevity we use $MAD^{OC}$ to indicate the internal velocity dispersion of both.}.
We only selected RVs with an error of less than 3.8~\kmsec\ (which is the median error) and stars with the Gaia flag {\it non\_single\_star = 0} in order to neglect measurements that inflate the RV distribution. The MAD error bars were computed by bootstrapping the sample and taking the standard deviation of the resulting MADs.

\begin{figure*}[t]
\centering
\includegraphics[width=16cm, clip]{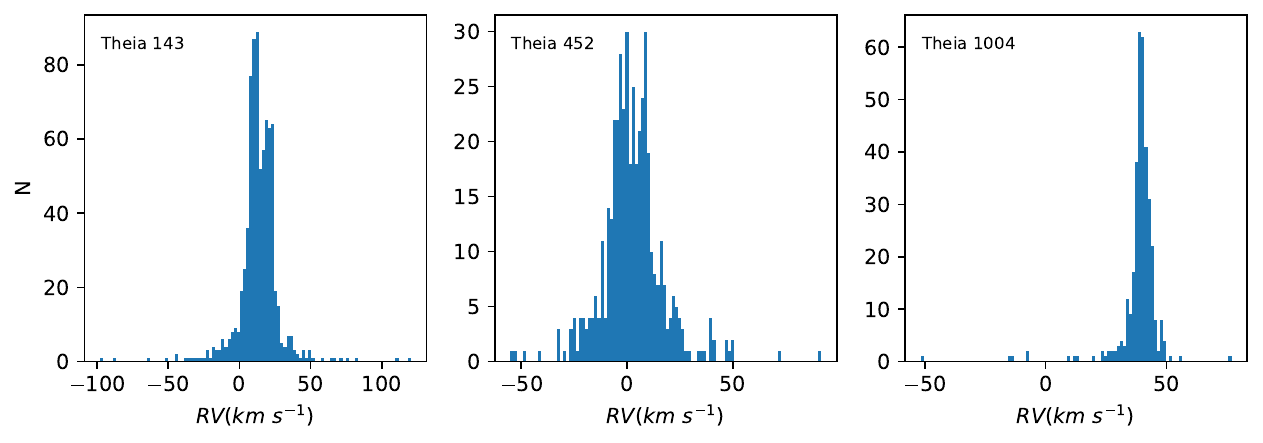}
\caption{Radial-velocity distributions of three objects, two of which (Theia~143 and Theia~452) show a bimodal distribution.} 
\label{RV_distrib}
\end{figure*}

The convenience of using the MAD over the standard deviation $\sigma$ to express the velocity dispersion is due to the fact that the MAD is less affected by outliers. Such outliers affect the RV distribution of the objects, enlarging $\sigma_{RV}$ to the extent that it is incompatible with $\sigma_{v_t}$.  
This disagreement was pointed out by \citet{zucker} (and was clearly visible in Fig.~12 of KC19), whose authors noted that while the tangential-velocity dispersions of the strings are mostly lower than $2.5$~\kmsec, their radial-velocity dispersion spans the 5--40~\kmsec\ interval with an average of RV$\sim16$~\kmsec. This discrepancy is due to the identification process of the string's members used by KC19, which employs the machine-learning algorithm HDBSCAN for the variable Galactic coordinates $l$, $b$, parallax $\pi$, and proper motions and neglects RVs. 
On the one hand, HDBSCAN looks for over-densities that must be limited in the velocity space in order to be found. Therefore, the found structure exhibits an internal velocity dispersion that represents the velocity constraints given by HDBSCAN to find the over-density. In other words, the internal velocity dispersion in proper motion cannot be wider than the quantity allowed by the method employed. 
On the other hand, neglecting the third velocity component allowed outliers in RVs to be misclassified as members. Such outliers inflate the internal RV dispersion to the observed level of $\sim16$~\kmsec.
We amended the situation by using the MAD that expresses the spread of the core of the distribution neglecting outliers and wide wings\footnote{We corroborated the robustness of the MAD by fitting the RV distributions with a Gaussian and inferring its standard deviation, $\sigma_g$. We found that the $\sigma_g$ is just slightly higher than the MAD (as it must be, since for a Gaussian distribution we have $MAD \simeq 0.8\sigma_g $), and both are much lower than the computed standard deviation.}.
Moreover, the analysis of the RVs distribution allowed us to find that Theia~452 and Theia~143 (two of the objects studied by us) show a bimodal RV distribution, indicating that these objects should be split in two parts each (see Fig.~\ref{RV_distrib}).

In the following, we find that the $MAD_{RV}$s are systematically higher than $MAD_{v_t}$s. To explain this difference, we recall that the RV uncertainty is dependent on the magnitude of the star, with errors below 1~\kmsec\ for magnitudes $G_{RVS}<10$ and errors greater than 2.5~\kmsec\ for $G_{RVS}>12$ \citep{sartoretti}. Given that the objects under study are not bright, we expect errors larger than 1~\kmsec. 
\citet{soubiran2018} studied the RVs of more than 1000 OCs from Gaia DR2 data, finding that the RV standard deviations of OCs  are in the range from 0.1 to 20~\kmsec,\ with most of them between 1 and 5~\kmsec\ and an average of 1.7~\kmsec.
Therefore, the higher $MAD_{RV}$s observed are the result of the sum in quadrature of the intrinsic internal velocity dispersion of the OCs and the errors in the RVs. Later, in Sect.~\ref{sec_validation}, we discuss the effect of such uncertainties.

\section{Validation}\label{sec_validation}
In order to validate our methods we chose a few objects associated with well-known OCs. Here we discuss the internal consistency of the
kinematics with the abundances expected for OC members compared
to the objects under investigation.

The Hyades member sample, according to \citet{meingast_hyades}, counts 237 stars. 
After matching them with Theia~1004 members we selected members with an RV error smaller than 3.8~\kmsec\ and Gaia flag {\it non\_single\_star} = 0, leaving us with a sample that counts 123 stars.
The $v_t$ internal velocity dispersion of this sample is 
$MAD_{v_t}^{OC} = 0.37\pm0.04$~\kmsec, which is in reasonable agreement with the literature \citep{oh, madsen}. On the other hand, the RV internal dispersion is $MAD_{RV}^{OC} = 1.28\pm0.15$~\kmsec.
For chemistry, we investigated the iron-abundance dispersion using the GALAH iron abundance [Fe/H]. We cross-matched the stars in Theia~1004 that have GALAH data with the \citet{meingast_hyades}
catalog and obtained 23 Hyades members with GALAH abundances. These stars have an iron distribution spread of
$\sigma_{[Fe/H]}^{OC} = 0.119\pm0.003$~dex, which looks slightly higher than
that expected of an OC. In fact, there are two stars that deviate more than
$2\sigma_{[Fe/H]}$ from the mean. These two stars---GaiaDR3~145664777917536512 and
GaiaDR3~3410453489023420416---have iron abundances that are 0.36~dex 
and 0.27~dex lower than the mean $\overline{[Fe/H]}_{Hyades} = 0.084\pm0.017$~dex.
They can therefore be considered non-members on the basis of the chemical
discrepancy and removed from the following abundance analysis. 
When such stars are removed, we find $\sigma_{[Fe/H]}^{OC} =
0.066\pm0.002$~dex, in agreement with the dispersion found by using the abundances of
Hyades members by \citet{takeda}.\\

The Pleiades counts 1245 members according to \citet{roser2020}.
Cross-matching this sample with Theia~369 and selecting stars 
with RV error smaller than 3.8~\kmsec\ and Gaia flag {\it non\_single\_star} = 0,
we find 258 stars in common with internal 
velocity dispersions of $MAD_{v_t}^{OC} = 0.36\pm0.02$~\kmsec\ \citep[in fair agreement with the literature; see][]{makarov} and $MAD_{RV}^{OC}=1.09\pm0.10$~\kmsec.
The latter is in very good agreement with the dispersion $\sigma_{RV}=1.36$\,\kmsec\ found by \citet{Frasca2025} if we take the ratio of 0.8 between $MAD$ and $\sigma$ into account.
Among these stars, GALAH reports abundances for 27 of them. The spread of their iron-abundance distribution is $\sigma_{[Fe/H]} =
0.071\pm0.001$~dex, in agreement with the dispersion of 0.06 dex reported by \citet{Frasca2025}.\\

NGC~2516 counts 1726 star members (with probability $>0.9$) according to \citet{tarricq}.
The cross-match with Theia~613 and the selection of stars 
with an RV error smaller than 3.8~\kmsec\ and the Gaia flag {\it non\_single\_star} = 0 leaves a sample of 422 stars, whose internal velocity distribution spreads are $MAD_{v_t}^{OC} = 0.53\pm0.03$~\kmsec\ and $MAD_{RV}^{OC} = 1.8\pm0.1$~\kmsec \citep[as expected for OCs; ][]{madsen, olivares}.
Among these stars, GALAH provides abundances
for 41 of them. The spread of their iron abundance distribution is $\sigma_{[Fe/H]} =
0.043\pm0.001$~dex, which is in agreement with what we expect from an OC.\\

The numbers reported above show that when we have
accurate OC member selection the internal velocity dispersions can be below 1~\kmsec\ for $v_t,$ as expected for OCs, and 1-2~\kmsec\ for RVs, which is in agreement with the literature \citep{soubiran2018}.
This gives us confidence in the velocity dispersion derived from the $v_t$ and RVs and tells us that we can trust the MADs we report in
Sect.~\ref{sec_results} for the objects as a whole.\\

Concerning the uncertainties affecting the $MAD$s, we note $MAD_{v_t}^{OC} < 1$~\kmsec\ for all three cases studied, while
their $MAD_{RV}^{OC}$s are higher, i.e., between 1 and 2~\kmsec. This discrepancy can be due to undetected binaries, but this is most likely due to measurement uncertainties.
In fact, while the tangential-velocity errors are in the median of $\sim0.2$~\kmsec, the RV errors are greater, given that we selected RVs with errors smaller than 3.8~\kmsec\ to compromise on accuracy and completeness. An RV error of $~\sim3$~\kmsec\ can in fact significantly inflate the measurements of $MAD_{RV}$ when the intrinsic RV dispersion is the same size or smaller. To test this possibility, we repeated the measurement of $MAD_{RV}^{OC}$s for the three above-mentioned OCs by selecting stars whose RV errors are smaller than 1.5~\kmsec. We found $MAD_{RV}^{OC}=1.3\pm0.15$, $0.88\pm0.08$, and $1.14\pm0.16$~\kmsec\ for Hyades, Pleiades, and NGC~2516, respectively. Comparing these values with those reported earlier in this section, we observe that for Hyades there is no significant change, while for Pleiades and NGC~2516 the $MAD_{RV}^{OC}$ is significantly lower and closer to the $MAD_{v_t}^{OC}$ values. We conclude that RV uncertainties can inflate the value of $MAD_{RV}$. In Sect.~\ref{sec_results} we test the effect of this uncertainty on the objects under study and evaluate its impact on the results.

\section{Results}\label{sec_results}
\begin{figure*}[t]
\sidecaption
\includegraphics[width=12cm, clip]{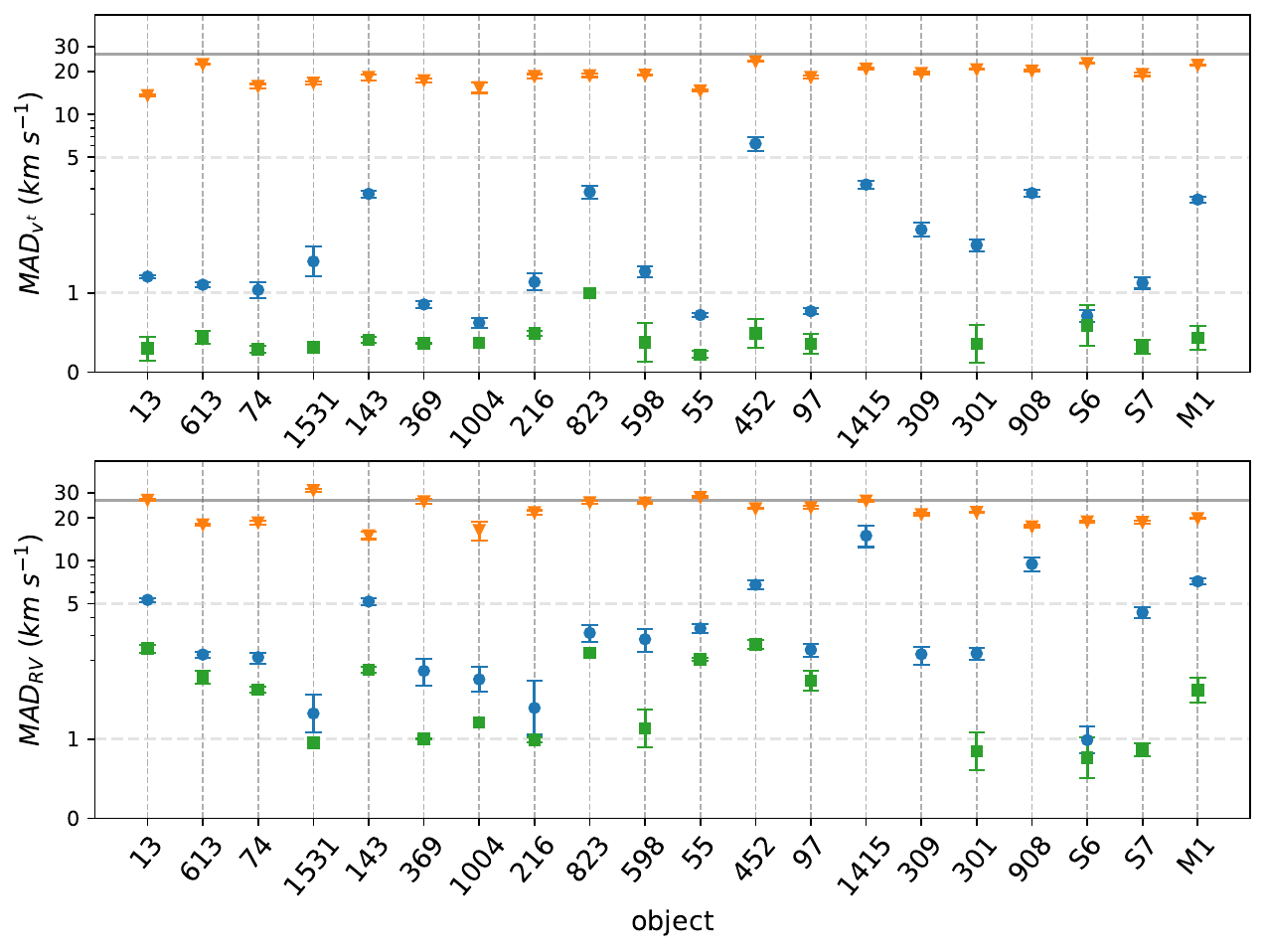}
\caption{Internal velocity dispersions $MAD_{v_t}$ (top panel) and $MAD_{RV}$ 
(bottom panel) for the objects
under investigation. Blue points represent the MAD of the objects
(excluding the stars belonging to OCs). Green squares represent the mean of
the MAD of the OCs associated with the object. Orange triangles
represent the MAD of the test samples divided by 0.8 in order to compare the value to the mean of the three dispersions $\sigma_R$, $\sigma_{\phi}$, and $\sigma_Z$ of the thin disk \citep[according to][]{borja},
which is indicated by the horizontal solid gray line.}
\label{MAD_PM_RV_objects}
\end{figure*}

As shown in Figs.~\ref{diagnostic1}, \ref{diagnostic2}, and \ref{diagnostic3},
most of the objects 
have one or more associated
OCs. We verified that a significant part of the OC members belong to the
associated object\footnote{Not all of them belonging to the
object may be due to the fact that this was sampled from $Gaia$ DR2
while the HR24 OCs were sampled from $Gaia$ DR3. In fact, when the OC
member identification comes from \citet{cantat-gaudin2020A} (which is based
on $Gaia$ DR2), all the OC members belong to the associated objects.} 
(see Table~\ref{tab:clusters} for the number of such members).
In Table~\ref{tab_int_vel} we report the calculated internal velocity
dispersion of the objects as a whole, while in Fig.~\ref{MAD_PM_RV_objects} 
we compare the internal velocity dispersions of the objects in $v_t$ and RV, 
their test samples, and the mean internal velocity dispersion of the OCs
associated with the objects.

\begin{table*}
\caption{Internal velocity dispersions of the objects and their test
samples for tangential velocities and RVs expressed in \kmsec.}
\small
\label{tab_int_vel}
\centering
\begin{tabular}{r|rrrrr|rrrrr||rr}
\hline\hline
Theia N & $MAD_{v_t}^{obj}$ & $N_{v_t}^{obj}$\tablefootmark{(a)} &  $MAD_{RV}^{obj}$ & $N_{RV}^{obj}$ & $ME_{RV}^{obj}$\tablefootmark{(b)} &  $MAD_{v_t}^{test}$ & $N_{v_t}^{test}$ &  $MAD_{RV}^{test}$ & $N_{RV}^{test}$ & $ME_{RV}^{test}$ &  $mad_{RV}^{obj}$\tablefootmark{(c)} & $n_{RV}^{obj}$\\
\hline
13 & 1.21$\pm$0.02 & 7990 & 5.31$\pm$0.17 & 1070 & 2.31 & 10.90$\pm$ 0.28 & 1070 & 21.42$\pm$ 0.54 & 2303 & 1.71 & 5.66$\pm$0.36 & 291\\
613 & 1.10$\pm$0.03 & 1797 & 2.19$\pm$0.11 & 751 & 2.31 & 18.02$\pm$ 0.27 & 751 & 14.38$\pm$ 0.23 & 4939 & 1.19 & 1.80$\pm$0.22 & 133\\
74 & 1.04$\pm$0.10 & 547 & 2.10$\pm$0.15 & 347 & 2.18 & 12.62$\pm$ 0.58 & 347 & 14.84$\pm$ 0.68 & 785 & 1.60 & 1.80$\pm$0.17 & 96\\
1531 & 1.40$\pm$0.19 & 125 & 1.33$\pm$0.23 & 120 & 1.06 & 13.32$\pm$ 0.44 & 120 & 25.03$\pm$ 0.83 & 1156 & 0.84 & 0.83$\pm$0.18 & 74\\
143 & 2.76$\pm$0.15 & 596 & 5.17$\pm$0.30 & 353 & 2.60 & 14.62$\pm$ 0.88 & 353 & 12.04$\pm$ 0.86 & 357 & 1.63 & 5.22$\pm$0.80 & 57\\
369 & 0.85$\pm$0.05 & 661 & 1.86$\pm$0.19 & 404 & 0.87 & 13.88$\pm$ 0.52 & 404 & 20.89$\pm$ 0.92 & 939 & 0.91 & 1.46$\pm$0.19 & 268\\
1004 & 0.62$\pm$0.06 & 171 & 1.76$\pm$0.15 & 264 & 0.48 & 12.39$\pm$ 1.35 & 264 & 13.08$\pm$ 2.45 & 141 & 0.36 & 1.43$\pm$0.15 & 175\\
216 & 1.14$\pm$0.11 & 218 & 1.40$\pm$0.34 & 121 & 0.94 & 14.84$\pm$ 0.66 & 121 & 17.46$\pm$ 0.80 & 912 & 0.41 & 1.17$\pm$0.18 & 86\\
823 & 2.85$\pm$0.29 & 140 & 3.11$\pm$0.43 & 154 & 2.05 & 15.02$\pm$ 0.46 & 154 & 20.56$\pm$ 0.56 & 1704 & 1.00 & 2.92$\pm$0.95 & 46\\
598 & 1.27$\pm$0.07 & 519 & 2.81$\pm$0.53 & 195 & 1.33 & 15.16$\pm$ 0.22 & 195 & 20.47$\pm$ 0.33 & 5707 & 0.84 & 1.52$\pm$0.35 & 107\\
55 & 0.72$\pm$0.03 & 946 & 3.35$\pm$0.26 & 304 & 2.17 & 11.75$\pm$ 0.29 & 304 & 22.39$\pm$ 0.48 & 2736 & 1.81 & 3.45$\pm$0.51 & 95\\
452 & 6.23$\pm$0.70 & 1007 & 6.78$\pm$0.50 & 228 & 2.38 & 18.85$\pm$ 0.20 & 228 & 18.63$\pm$ 0.21 & 11454 & 1.23 & 9.73$\pm$2.34 & 48\\
97 & 0.77$\pm$0.04 & 458 & 2.37$\pm$0.24 & 230 & 2.25 & 14.59$\pm$ 0.40 & 230 & 18.98$\pm$ 0.60 & 1774 & 1.47 & 1.89$\pm$0.32 & 46\\
1415 & 3.21$\pm$0.21 & 258 & 15.0$\pm$2.52 & 57 & 2.10 & 16.73$\pm$ 0.28 & 57 & 21.15$\pm$ 0.37 & 4001 & 1.52 & 9.87$\pm$4.82 & 16\\
309 & 1.80$\pm$0.09 & 570 & 2.20$\pm$0.27 & 156 & 1.45 & 15.59$\pm$ 0.36 & 156 & 16.95$\pm$ 0.36 & 3257 & 0.65 & 1.83$\pm$0.27 & 81\\
301 & 1.61$\pm$0.07 & 1003 & 2.23$\pm$0.21 & 377 & 0.93 & 16.57$\pm$ 0.18 & 377 & 17.54$\pm$ 0.20 & 13219 & 1.50 & 1.97$\pm$0.21 & 235\\
908 & 2.80$\pm$0.15 & 519 & 9.48$\pm$1.07 & 167 & 0.97 & 16.22$\pm$ 0.37 & 167 & 13.91$\pm$ 0.38 & 3138 & 0.86 & 9.07$\pm$1.40 & 108\\
S6 & 0.71$\pm$0.08 & 97 & 0.99$\pm$0.17 & 62 & 0.55 & 18.30$\pm$ 0.26 & 62 & 15.08$\pm$ 0.23 & 7310 & 0.66 & 0.80$\pm$0.16 & 47\\
S7 & 1.13$\pm$0.07 & 516 & 4.34$\pm$0.37 & 153 & 1.03 & 15.31$\pm$ 0.55 & 153 & 14.93$\pm$ 0.45 & 1382 & 0.74 & 3.96$\pm$0.51 & 102\\
M1 & 2.52$\pm$0.11 & 922 & 7.16$\pm$0.32 & 767 & 0.76 & 17.79$\pm$ 0.28 & 767 & 15.90$\pm$ 0.29 & 5120 & 0.69 & 7.03$\pm$0.37 & 522\\
\hline
\end{tabular}
\tablefoot{
\tablefoottext{a}{$N$ is the number of stars
used to compute the $MAD$ in tangential velocities and RVs for the different samples.}
\tablefoottext{b}{$ME_{RV}^{obj}$ and $ME_{RV}^{test}$ represent the median RV errors of the objects and test samples.}
\tablefoottext{c}{The median absolute deviation, $mad_{RV}^{obj}$, and the number of stars, $n_{RV}^{obj}$, were derived by choosing only stars with RV errors lower than 1.5~\kmsec\  in order to test the robustness of the results, as discussed in Sect.~\ref{sec_validation} and Sect.~\ref{sec_results}.}

}
\end{table*} 

For the $v_t$ (top of Fig.~\ref{MAD_PM_RV_objects}) these three populations occupy three different velocity ranges. While the $\overline{MAD}_{v_t}^{OC}$s  are mostly in the 0--1~\kmsec\ range, the $MAD_{v_t}^{obj}$s are always higher and
mostly in the 1--5~\kmsec\ range, while the internal dispersions of the test samples
are close to the Galactic thin-disk velocity dispersion, 26.4~\kmsec\ , computed by \citet{borja} as the  mean of the three velocities $\sigma_R$, $\sigma_{\phi}$, and $\sigma_z$.
For RVs (bottom of Fig.~\ref{MAD_PM_RV_objects}) the $MAD_{RV}^{obj}$s mainly lie in the 1--7~\kmsec\ range (excluding Theia~908 and Theia~1415; see Sects.~\ref{sec908} and ~\ref{sec1415}),
and $\overline{MAD}_{RV}^{OC}$s lie in the 1--5~\kmsec range, with $MAD_{RV}^{obj}$s being higher than $\overline{MAD}_{RV}^{OC}$s on average. The internal dispersions of the test samples are in good agreement with \citet{borja}. 
As pointed out in Sect.~\ref{sec_validation} the higher $MAD_{RV}$s values observed with respect to the $MAD_{v_t}$s ones are due to the errors of the Gaia RVs, which tend to inflate the RV dispersions of OCs and objects as well. 

On the right side of Table~\ref{tab_int_vel} we report the RV median absolute deviation and the sample size when only stars with RV errors smaller than 1.5~\kmsec\ are selected. We call this $mad_{RV}^{obj}$ (lower case) to distinguish it from the working samples chosen with an RV error lower than 3.8~\kmsec\ indicated with $MAD_{RV}^{obj}$ (upper case, explained in Sect.~\ref{sec_kinematic}).
We observe that for samples with a lower RV error threshold, just four objects (Theia~1531, Theia~369, Theia~598, Theia~452) have $mad_{RV}^{obj}$ significantly lower (at the 1$\sigma$ level) than $MAD_{RV}^{obj}$, while for most objects $mad_{RV}^{obj}$ and $MAD_{RV}^{obj}$ agree inside the uncertainties with just a general tendency for the former to be slightly lower than the latter. This
 set a lower limit to the detection of the lower velocity dispersions (as discussed in Sect.~\ref{sec_kinematic}). Therefore, we must assume that the velocity dispersions reported at the bottom of Fig.~\ref{MAD_PM_RV_objects} are the upper limits because they are the result of the sum of the internal velocity dispersion and the RV errors. However, such limits do not significantly affect the RV dispersion of the objects, which are in both cases found in the 1--7~\kmsec\ range.

\subsection{Object features}
In the following, we briefly discuss each object one by one. Here,
we preserve the distinction between OCs and MGs of HR24, although we do not use it throughout the paper.
We report the internal velocity dispersion $MAD_{RV}$ only, keeping in mind that it is an upper limit. For a complete comparison between $MAD_{v_t}$ and $MAD_{RV}$ we refer the reader to Table~\ref{tab_int_vel}.

\subsubsection{Theia~13}

This object is part of the Orion association. It shows
substructures in (X,Y) and velocity space and embeds 
27 star clumps classified as OCs or MGs by HR24. The
GALAH stars seem to be fairly distributed; therefore, the chemical abundances are representative of the whole string. Their general
abundance distribution is narrow ($\sigma_{[Fe/H]}=0.085\pm0.007$~dex). The
internal RV dispersions of the single OCs are $\leq2$~\kmsec\ 
(see Table.~\ref{tab:clusters} for the precise values), while for the whole string it is 
$MAD_{RV}^{obj} = 5.3\pm0.2$\kmsec.
It is a complex structure for which the investigation goes beyond the data available in this work,
and here we only report a few global features that we observed.\\

\subsubsection{Theia~613}
This object embeds the OCs NGC~2516 (at its center) and HSC~2336.
The GALAH stars are located mostly close to the center of
NGC~2516; therefore, their abundances are representative of this
cluster and not of the whole string. In fact, the abundances have
dispersions that are narrower than those of the test sample, as
expected from an OC. The internal velocity dispersion of the object is
$MAD_{RV}^{obj} = 2.2\pm0.1$\kmsec.
\\

\subsubsection{Theia~74}
According to KC19, Theia~74 is associated with NGC~2547, but we also found
five other associated OCs (see Table~\ref{tab:clusters}) and one MG. 
They can be identified in the cartesian (X, Y) and Galactic coordinate
planes as elongated blobs of stars (see Fig.~\ref{diagnostic1}). 
In particular, the GALAH stars that belong to Theia~74 are members
of the Collinder~135, Collinder~140, and NGC~2451B clusters.
Therefore, the chemical information is only representative of these 
three clusters. From Figs.~\ref{fig_mean_std} and 
\ref{fig_XH_distr_mean_std_diff} we see that the distributions of the
chemical abundances are narrow (i.e., chemically homogeneous, as we expect 
for OCs) and clearly distinct from its test sample (Fig.~\ref{fig:tt-values}). 
\citet{swiggum} reported that these three OCs belong to a ``family''
(which they called the Cr135 family) of 39 different OCs. By integrating the orbits backward, they found that these clusters reached a minimum size configuration $21.7^{+10.5}_{-8.6}$~Myr ago (corresponding to $7.34^{+0.17}_{-0.20}$ in $\log_{10}$). We investigated the chemistry of the three
OCs, since they have a reasonable number of GALAH stars each (12 for
Collinder~135, 9 for Collinder 140, and 21 for NGC~2451B). 
We computed the p-values of the 
t-test among the three clusters in their three combinations. We find that they are 
chemically indistinguishable from each other for all elements
considered, with p-values all higher than 0.05.
They also have very similar ages, 
i.e., $\log Age =$ 7.42, 7.43, and 7.61
for Collinder~135, Collinder~140, and NGC~2451B, respectively
\citep{cantat-gaudin2020B}, which are in agreement with the Swiggum minimum size configuration time. We can conclude that these three clusters must have been born together, and this is confirmed by their chemical similarity. Thus, it is not surprising to find that
the general standard deviation is as narrow as $\sigma_{[Fe/H]} =
0.055\pm0.005$~dex.  The internal velocity dispersions $MAD_{RV}^{OC}$ 
of the three OCs are between 1.2 and 3.4~\kmsec, while the overall
object has $MAD_{RV}^{obj} = 2.1\pm0.1$~\kmsec.\\
This string also matches one of the populations (in particular the fourth) identified by
\citet{cantat-gaudin2019} in the Vela-Puppis region. In fact, the authors reported that
the OCs Collinder~135, Collinder~140, NGC~2451B, NGC~2547, and Alessi~36
(alias UBC~7) are physically connected by a continuous distribution of stars.
In addition, this structure belongs to the western  part of the snakes identified by
\citet{wang} together with Theia~143 (while Theia~55 belongs to the eastern part).\\

\subsubsection{Theia~1531}
It embeds only the OC Ruprecht~147 and the stars observed by GALAH
clump around its center. The low internal velocity dispersion and the
narrow chemical-abundance distribution are representative of the OC.
KC19 classified it as a compact and roundish object, although in
Fig.~\ref{diagnostic1} it is elongated and exhibits tails that may be
of tidal origin. This is supported by \citet{tarricq}, whose authors found a tidal tail size of $16\pm4$~pc compared to a core radius of $1.5\pm0.6$~pc.
The internal velocity dispersion of the object is
$MAD_{RV}^{obj}=1.3\pm0.2$~\kmsec, which is slightly higher than that of the OC Ruprecht~147, i.e., $MAD_{RV}^{OC}=0.9\pm0.2$~\kmsec.
We highlight the position of this object, which lies on the metal-rich side of
Fig.~\ref{fig_mean_std}, and the  Levene's test and t-test p-values in Fig.~\ref{fig:tt-values},
which establishes that the object is chemically distinct from the local field stars.

\subsubsection{Theia~143}
According to KC19 it
is associated with Trumpler~10. We find that four additional OCs and one MG are 
also embedded in it: Alessi~5, ASCC~58, BH~99, CWNU~287, and HSC~2215. 
These clusters appear as clumps of stars in
spatial and velocity spaces connected by a bridge of stars.
The GALAH stars focus mostly on Trumpler~10, while only one and two stars are
located close to Alessi~5 and BH~99, respectively. 
The GALAH star sample is therefore
representative of Trumpler~10 chemistry, which in fact appears
homogeneous and distinct from the field stars, given that all
elements have a  Levene's test p value below 0.05. 
The internal velocity dispersion of Trumpler~10 is consistent with an OC with $MAD_{RV}^{OC} =
1.6\pm0.3$~\kmsec, while the whole object has $MAD_{RV}^{obj} =
5.1\pm0.3$~\kmsec.  However, it must be noted that the RV distribution is bimodal (see Fig.~\ref{RV_distrib}). This suggests that this string is actually made up of two parts with different RVs. 
Looking at embedded OCs, we find that Trumpler~10 and MG HSC~2215 have RVs of 18.7 and 22.3~\kmsec\ (respectively), while the other four OCs have RVs spanning from 8.4 to 10.9~\kmsec. These RVs match the bimodality of the RV distribution of the string seen in Fig.~\ref{RV_distrib}, where two peaks are found at 20.4 and 9.5~\kmsec. This confirms that Theia~143 is actually two distinct structures.
Theia~143 is located in the Vela-Puppis region, 
and its members partially overlap the second population described by 
\citet[][]{cantat-gaudin2019}, with 546 stars in common out of 1759. 
Theia~143 is located close to Theia~74, and it may be
interesting to compare the chemistry of Trumpler~10 with the three OCs Collinder~135, 
Collinder~140, and NGC~2451B embedded in Theia~74. In
Table~\ref{tab:trump10_pvalues} we report the p-values of the t-test between
Trumpler~10 and Collinder~135, Collinder~140, and NGC~2451B for each element
abundance.

\begin{table}[!th]
\caption{T-test p-values computed for each element abundance distribution of
Trumpler~10 versus Collinder~135, Collinder~140, and NGC~2451B. Values
smaller than 0.05 are highlighted in bold.}
\label{tab:trump10_pvalues}
\centering
\begin{tabular}{l|rrr}
\hline\hline 
[El/H] & \multicolumn{3}{c}{Trumpler~10 vs.} \\
       & Collinder~135 & Collinder~140 & NGC~2451B\\
\hline
Fe & {\bf 0.04} & 0.17 & {\bf 0.00} \\
Na & 0.69 & 0.74 & 0.21 \\
Mg & 0.58 & 0.33 & 0.25 \\
Si & 0.18 & 0.12 & 0.10 \\
Ca & 0.13 & 0.07 & 0.19 \\
Sc & 0.41 & 0.49 & 0.09 \\
Ti & {\bf 0.01} & 0.08 & {\bf 0.01} \\
Cr & {\bf 0.02} & 0.11 & {\bf 0.00} \\
Mn & 0.07 & 0.14 & 0.10 \\
Co & {\bf 0.00} & 0.12 & {\bf 0.01} \\
Ni & {\bf 0.04} & 0.11 & {\bf 0.00} \\
\hline
\end{tabular}
\end{table}

While Collinder~140 appears chemically indistinguishable from Trumpler~10,
Collinder~135 and NGC~2451B have five elements with p-values of $<0.05$, which
means that they are statistically distinct from Trumpler~10. More precisely, Trumpler~10 is
slightly more abundant than Collinder~135 and NGC~2451B in all elements.
All these OCs being young \citep[Trumpler~10 has $\log Age = 7.51$ according to ][]{cantat-gaudin2020B},
their chemistry may represent the local field, suggesting that the volumes covered by Theia~74 and Theia~143 may have a
different chemical fingerprint. In fact, we note that according to \citet{swiggum}, Trumpler~10 does not belong to the Cr135 family but to the m6 family, which is older. As reported before, this object is part of the west side of the ``snake'' identified by \citet{wang}.
\\

\subsubsection{Theia~369}
The Pleiades (Melotte~22) are at the center of this roundish (non-string-like) object with GALAH targeting its members. However, this object embeds
two other MGs: HSC~1580 and UPK~303. The chemical abundance 
distributions are narrow, as expected for an OC, and the internal velocity
dispersion is $MAD_{RV}^{OC}=1.0\pm0.1$~\kmsec. We have no RVs for HSC~1580 and UPK~303. The object as a whole has
$MAD_{RV}^{obj}=1.8\pm0.2$~\kmsec, which is higher than
the embedded OC. This can be explained if the object also embeds the
coronae or tidal tails that the Pleiades may have \citep{meingast2021, Frasca2025}, or considering that the object embeds two MGs that can inflate the internal velocity dispersion.
We state in Sect.~\ref{sec_validation} that it has a low iron-abundance
dispersion ($\sigma_{[Fe/H]}=0.071\pm0.001$~dex) when the \citet{roser}
membership is used. When using the HR24 membership instead, we have
$\sigma_{[Fe/H]}=0.09\pm0.01$~dex.

\subsubsection{Theia~1004}
This object embeds the Hyades OC. We state in Sect.~\ref{sec_validation} that it
has low internal velocity dispersion ($MAD_{v_t}^{OC}=0.54\pm0.02$~\kmsec\ and $MAD_{RV}^{OC} = 1.3\pm0.1$~\kmsec) 
and narrow abundance distributions. This must be compared with the
significantly higher internal velocity dispersion of the object, which is
$MAD_{RV}^{obj}=1.8\pm0.1$~\kmsec.  This difference can be explained 
if the object also embeds the tidal tails that the Hyades have \citep{meingast_hyades}.

\subsubsection{Theia~216}
According to KC19, Theia~216
is not associated with any OC, but according to HR24,
UPK~612 and HST~2712 are embedded in it and are classified 
as MGs\footnote{\citet{qin} classifies UPK~612 as OC.}.  
Their internal velocity dispersions are
$MAD_{RV}^{OC}=0.7\pm0.2$ and $MAD_{RV}^{OC}=1.3\pm0.8$~\kmsec, respectively, while the entire string has
$MAD_{RV}^{obj}=1.5\pm0.3$~\kmsec.
Considering that its extension in Galactic $l$ is about 20 degrees and that
the estimated radius containing half of the members is about 2.5 degrees
\citep{cantat-gaudin2020A}, from Fig.~\ref{diagnostic1}, we infer that
the GALAH stars probe the whole
structure and do not focus on the center of the cluster. 
It is interesting to note that the chemical-abundance dispersion
representative of the whole string is tight ($\sigma_{[Fe/H]}=0.04\pm0.006$~dex); it is
compatible with an OC and clearly chemically distinct from its test sample
(all elements have Levene's test p-values of $<0.05$). Besides this, it 
shows hints of chemical gradients (see Sect.~\ref{sec_216} for a discussion).\\

\subsubsection{Theia~823}
This elongated structure embeds the OC Mamajek~4. The
cluster has an internal velocity dispersion of $MAD_{RV}^{OC}=2.2\pm0.2$~\kmsec,\
while for the whole structure this is $MAD_{RV}^{obj}=3.2\pm0.5$~\kmsec.
The GALAH stars are distributed along the string such that the tight
abundance distributions ($\sigma_{[Fe/H]}=0.08\pm0.01$~dex) are
representative of the whole structure, which exhibits the chemical homogeneity of an
OC. It is chemically distinct from its test sample, with all elements 
having Levene's test p-values of $<0.05$.\\

\subsubsection{Theia~598}
This string embeds the OC Alessi~9 and
the MG HSC~2773. Their internal velocity dispersions are
$MAD_{RV}^{OC}=0.8\pm0.2$~\kmsec\
and $MAD_{RV}^{OC}=1.3\pm3.0$~\kmsec, respectively\footnote{The error bar for the MG HSC~2773
is larger than the measurement, which is due to the bootstrap method we used to measure the error bar;
therefore, this value is highly uncertain.}, whereas the whole
object has $MAD_{RV}^{obj}=2.8\pm0.5$~\kmsec. The GALAH stars are
distributed along the structure; therefore, the chemical abundance
distributions are representative of the whole string. The distribution in
[Fe/H] has $\sigma_{[Fe/H]}=0.10\pm0.01$~dex and is distinct from its
test sample, with the p-values of all elements of the Levene's test (but one) being $<0.05$.\\

\subsubsection{Theia~55}
It is an elongated structure with several blobs of
stars visible on the space and velocity planes that match the positions 
of three OCs and six MGs. Their $MAD_{RV}^{OC}$s span from 1.5 to 3.0~\kmsec,\
while $MAD_{RV}^{obj}=2.8\pm0.5$~\kmsec. The GALAH stars are
distributed along the structure; therefore, the chemical-abundance
distributions are representative of the whole string. It is chemically
distinct from the field stars because the Levene p-values are 
below 0.05 for most of the elements analyzed. As reported before, this object is part of the east side of the stellar structure (named ``the snake") identified by \citet{wang}.\\

\subsubsection{Theia~452}
It is made of two clouds of stars separated in spatial and velocity
spaces. The western part is associated with Alessi~24, while the other embeds
three MGs. Alessi~24 and CWNU~61 have 
$MAD_{RV}^{OC}=2.1\pm1.1$ and $2.7\pm1.8$~\kmsec, respectively, and we have no measurements for the other two MGs. The whole
object has $MAD_{RV}^{obj}=6.7\pm0.5$~\kmsec. The RV distribution exhibits a double peak at $RV\sim-3$ and $7$~\kmsec, suggesting that this object is composed of two distinct parts.
The GALAH stars are distributed along the string and show a fairly homogeneous chemistry with
$\sigma_{[Fe/H]}=0.09\pm0.02$~dex, but it has 
just four elements with Levene's test p-values of $\leq 0.05$.
This leaves us with no clear conclusion about the difference between Theia~452 and its test sample.\\

\subsubsection{Theia~301}
It embeds five groups classified as MGs. We have internal
velocity dispersions, $MAD_{RV}^{OC}$, of only three of them, which span from 0.7~\kmsec\ to
1.7~\kmsec,\ while the entire string has $MAD_{RV}^{obj}=2.1\pm0.3$~\kmsec.
The ten stars observed by GALAH provide average abundances similar to those of the
test sample, while the Levene's test p-values show that the dispersions of such
abundance distributions are distinct from the field stars for all but one
element (see Fig.~\ref{fig:tt-values}).
\\

\subsubsection{Theia~97}
It shows substructures in the space planes as well as
in the velocity plane. It embeds one OC and four MGs with $MAD_{RV}$ spanning from 0.6 to 2.6~\kmsec, while the dispersion of the whole string is $MAD_{RV}^{obj}=2.4\pm0.2$~\kmsec. It is chemically distinct from the
test sample because nine out of 11 elements have Levene's test p-values lower than
0.05.\\

\subsubsection{Theia~908}\label{sec908}
It is not associated with any known OC. Its overall internal
velocity dispersion is $MAD_{RV}^{obj}=9.5\pm1.0$~\kmsec,\ 
and the means and dispersions of its
elemental abundances are not distinguishable from the local field stars.
This leads us to think that this string may be a moving group of dynamical origin (see Sect.~\ref{sec_discussion}).\\

\subsubsection{Theia~309}
It is not associated with any known
OC. The internal velocity dispersion is $MAD_{RV}^{obj}=2.1\pm0.3$~\kmsec,\
while the dispersions of the elemental abundance distributions are tight similarly to an OC, with 
$\sigma_{[Fe/H]}=0.05\pm0.01$~dex and ten elements (out
of 11) with Levene p-values lower than 0.05.
It is interesting to compare the chemical abundances of Theia~309 and
Theia~216 because the two are spatially close to each other. 
Theia~309 is as chemically homogeneous as Theia~216. 
We performed the t-test and the Levene's test on the 11 elements considered between the 26 stars 
of Theia~216 and the ten stars of Theia~309.
For all elements, the p-values of the t-tests are well above 0.05, while for the
Levene's test the p-values are greater than 0.05 for all elements but one
(Ni). This tells us that the two
strings appear homogeneous and chemically indistinguishable.\\

\subsubsection{Theia~1415}\label{sec1415}
This elongated string is not associated with any known
OC. Its internal velocity dispersion is $MAD_{RV}^{obj}=15.2\pm2.4$~\kmsec\
while its chemistry is different from most of the other objects. 
In fact, it distinguishes itself by having a mean Fe abundance of $\overline{[Fe/H]}=0.076$~dex, much higher
than the other objects that have a mean Fe abundance around $-0.05$~dex. These
features suggest that it may be a moving group migrated from the inner disk (see
the discussion in Sect.~\ref{sec_discussion}).\\

\subsubsection{S6}
It holds the two MGs HSC~2407 and HSC~2705 with
$MAD_{RV}^{OC}=0.9\pm0.3$ and $0.6\pm0.5$~\kmsec, respectively.
The general internal velocity dispersion is 
$MAD_{RV}^{obj}=1.0\pm0.2$~\kmsec\ (R25 reports two internal velocity dispersions of $\sigma_{3D}=2.1$ and 2.7~\kmsec, depending on the method used to derive these values) and the elemental abundance
dispersions are narrow ($\sigma_{[Fe/H]}=0.08\pm0.02$~dex) with nine out of 11
elements with Levene's test p-values lower than 0.05. 
The GALAH stars probe the whole
structure and are therefore representative of the object chemistry.
Thus, this string is chemically distinct from its test sample.\\

\subsubsection{S7}
It embeds an MG (HSC~1849) and an OC (NGC~2451A), which have
$MAD_{RV}^{OC}=0.8\pm0.4$ and $1.0\pm0.2$~\kmsec, respectively. The string's internal velocity dispersion is 
$MAD_{RV}^{obj}=4.3\pm0.4$~\kmsec\ (R25 reports two internal velocity dispersions of $\sigma_{3D}=3.5$ and 5.7~\kmsec, depending on the method used to derive these values). The GALAH stars probe the whole
structure and are therefore representative of the object chemistry.
It is chemically distinct from its
test sample since nine out of 11 have Levene's test p-values below 0.05.
This string is chemically distinct from its test sample.\\

\subsubsection{M1}
Meingast~1 embeds 12 MGs  whose $MAD_{RV}^{OC}$s range from 0.6 to 2.6~\kmsec. The string's internal velocity dispersion is 
$MAD_{RV}^{obj}=7.2\pm0.3$~\kmsec. This value is not far from the $MAD_{RV}=5.0$~\kmsec\ obtained with the 42 RVs measured by \citet{hawkins}.
The GALAH stars are distributed along the whole
structure and are therefore representative of the object chemistry.
M1 is chemically homogeneous, with 
$\sigma_{[Fe/H]}=0.07\pm0.01$~dex, and distinct from its
test sample since nine elements out of 11 have Levene's test p-values below 0.05.\\

\subsection{Comparison with \citet{manea}}
\citet{manea} studied 18 Theia objects, ten of which are classified as
strings and the rest as groups. Using GALAH DR3 data \citep{galahDR3}
the authors investigated the chemical homogeneity with respect to open clusters and
nearby field stars. They found that some appear to be chemically
homogeneous, with tight abundance dispersions, while others have broader
dispersions. In Table~\ref{table_manea} we compare 
our results with theirs. 
We have five strings in common. For three of them (Theia 216, 309, and 1415) our conclusions are the same, while
for Theia~97 we found a disagreement with Manea's work.
In our results, its distribution in [Fe/H] is
$\sigma_{[Fe/H]}=0.07$~dex, which is narrow enough to claim its homogeneity, 
instead of $\sigma_{[Fe/H]}=0.17$~dex found by Manea and collaborators.
The difference found compared to the result of \citet{manea} may depend on the release of the GALAH catalog used: Manea et al. used DR3 and we used DR4.
For the remaining one, Theia~452, the result is not clear. The dispersion of the iron abundance is 
$\sigma_{[Fe/H]}=0.09\pm0.02$~dex, i.e., just at the limit of what we
consider homogeneous ($\sigma_{[Fe/H]}\leq0.1$~dex). Only four elements have
Levene's test p-values of $<0.05$, suggesting that the distinction from its test
sample is not clear.\\
Our conclusion about the homogeneity of the objects is in
fair agreement with \citet{manea}.

\begin{table}[h!]
\caption{Comparison of the chemical homogeneity of the Theia objects 
as derived by us and by \citet{manea}.} 
\label{table_manea}
\centering
\begin{tabular}{l|ll}
\hline\hline 
Theia& \multicolumn{2}{c}{is it homogeneous?} \\
\hline
    & this work & Manea et al.~2022\\
\hline
97 & yes  & no \\
216 & yes & yes \\
309 & yes & yes\\
452 & yes? & no \\
1415 & no & no \\
\hline
\end{tabular}
\end{table}

\subsection{Gradients of Theia~216}\label{sec_216}
\begin{figure*}[t]
\centering
\includegraphics[width=16cm, clip]{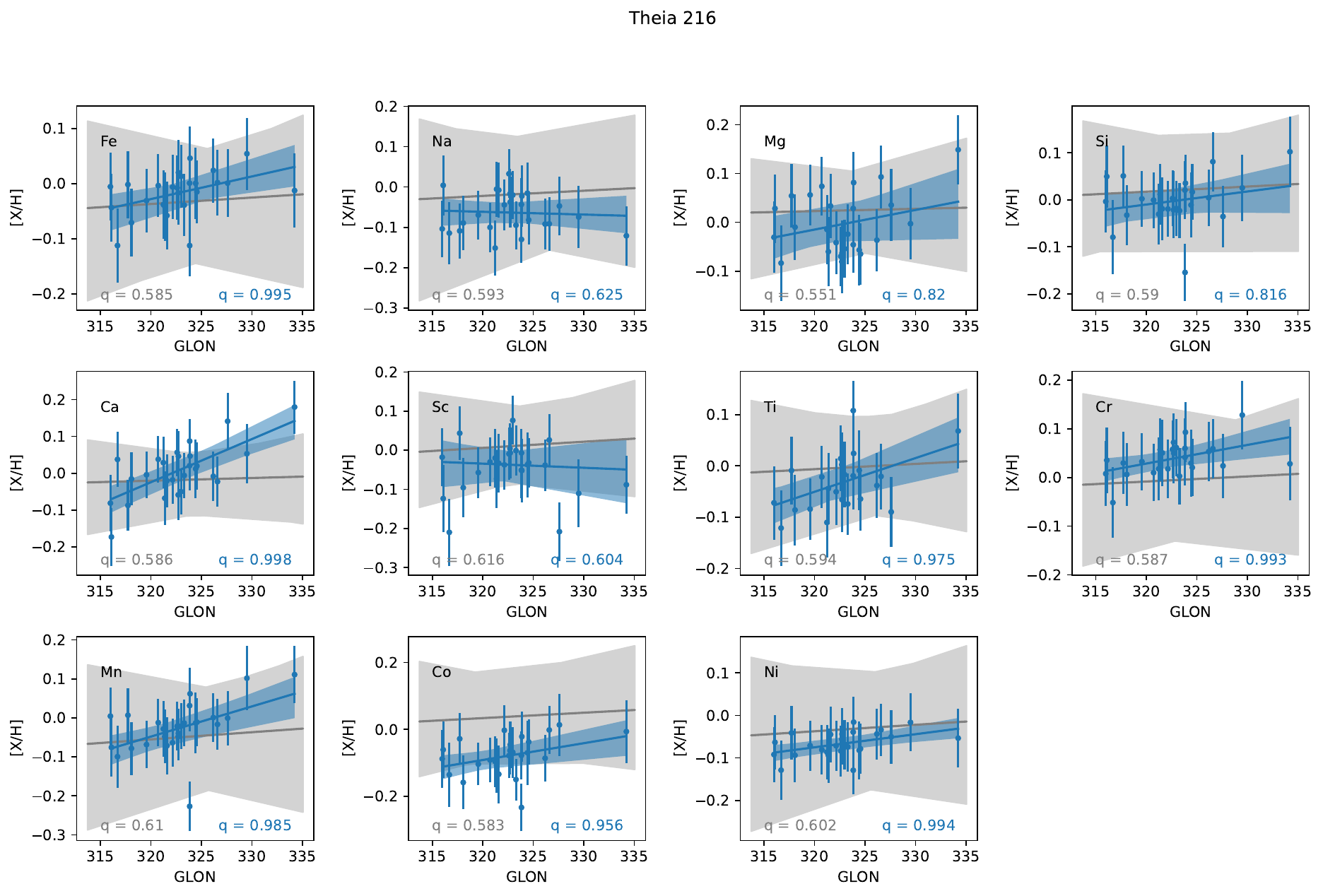}
\caption{Chemical gradients (blue lines) of Theia~216 (blue points). Blue shades show the uncertainty intervals computed with a bootstrap with resampling.  The same method was
applied to its test sample (gray lines and shading) for comparison purposes.
The ``q'' value expresses the quantile probabilities of the gradients.}
\label{fig_gradients_Theia216}
\end{figure*}

We find that Theia~216 is associated with UPK~612 and HST~2712.
It was classified as string by KC19, but it exhibits the chemical and
kinematic features 
of an OC: low internal velocity dispersion,
chemical homogeneity, and distinction from the surrounding field stars.
In particular, it shows clues of chemical gradients for several elements.
This may not be an exception, given that clues of chemical gradients have
been reported for Theia~456 by \citet{andrews}, whose authors found a gradient in
[Si/Fe] abundance. 
We measured the chemical-abundance gradients for the 11 elements considered
along the Galactic longitude, $l$. Theia~216 is in fact elongated in $l$ and
spans around 20 degrees. Gradients were measured with a weighted linear
regression performed on 1000 bootstrap resampling (with replacements) to estimate
the errors due to the distribution of the stars along $l$. Similarly, we
carried out the same analysis on its test sample in order to check the results
and errors on a field star sample of equal size. The results are shown
in Table~\ref{tab_grads_216} and displayed in
Fig.~\ref{fig_gradients_Theia216}. 

\begin{table}[!th]
\caption{Chemical gradients of Theia~216, its test sample, and the probability of having the sign reported 
for the 11 elements considered. The gradients and their standard deviations
are expressed in dex~deg$^{-1}$. The probabilities that are beyond 97.7\% ($2\sigma$) are in bold.} 
\label{tab_grads_216}
\centering
\begin{tabular}{l|rrrrrr}
\hline\hline 
ele & grad & $\sigma$ & prob & grad$_{test}$ & $\sigma_{test}$ & prob$_{test}$\\ 
\hline
Fe & 0.004 & 0.002 & \bf{0.995} & 0.001 & 0.006 & 0.585\\
Na & -0.001 & 0.002 & 0.625 &0.001 & 0.008 & 0.593\\
Mg & 0.004 & 0.004 & 0.820 & 0.000 & 0.005 & 0.551\\
Si & 0.003 & 0.003 & 0.816 &0.001 & 0.005 & 0.590\\
Ca & 0.012 & 0.003 & {\bf 0.998} & 0.001 & 0.005 & 0.589\\
Sc & -0.001 & 0.004 & 0.604 & 0.001 & 0.006 & 0.616\\
Ti & 0.007 & 0.003 & 0.975 & 0.001 & 0.005 & 0.594\\
Cr & 0.004 & 0.002 & {\bf 0.993} &0.000 & 0.006 & 0.587\\
Mn & 0.008 & 0.003 & {\bf 0.985} &0.001 & 0.008 & 0.610\\
Co & 0.005 & 0.003 & 0.956 &0.001 & 0.007 & 0.583\\
Ni & 0.003 & 0.001 & {\bf 0.994} &0.001 & 0.007 & 0.602\\
\hline
\end{tabular}
\end{table}

For Theia~216 the probability significance of the
gradients represented by the quantile values shows that for the element Ca
the gradient is positive with more than $3\sigma$ significance level.
At $2\sigma$ significance we find positive
gradients for the elements Fe, Cr, Mn, and Ni. Other elements show positive 
gradients of lower significance. In contrast, the test sample
shows no significant gradient in any element.

Given that the gradients found appear
statistically significant, we need
to discuss whether they are  physically meaningful.
It is known that the Galaxy has a chemical abundance variation along its
radius as well as its azimuth. 
The Galaxy radial mean gradient at the solar radius is
$\frac{d[M/H]}{R_G} \sim -0.06$~dex~kpc$^{-1}$
\citep[see, e.g.,][based on OCs, and references therein]{magrini}.
\citet{barbillon} also found azimuthal gradients revealing a patchy
abundance map in the solar neighborhood with 
metallicity gradients  that
can be estimated (by eye from their Fig.~4) on the order of 0.1~dex 
on the scale of few hundred parsecs in some locations of the Galactic plane.
For comparison purposes, we transformed the
chemical gradients expressed in dex~deg$^{-1}$ to
\dexkpc\ as follows:\\
$$
\frac{d[X/H]}{dL} = \frac{d[X/H]}{dl} \cdot \frac{\Delta l}{L},
$$
where $l$ is the Galactic longitude, $\Delta l$ is the angular extension of
the object along $l$, and $L$ is the linear length in kpcs of the string, 
as given in KC19.  The values, reported in Table~\ref{tab_grad_kpc_summary}, 
appear strikingly high, 
i.e., in the range of 1-2~\dexkpc,\ which is much higher than the Galactic radial gradient
and three to four times larger than the local azimuthal gradients found by \citet{barbillon}.

\begin{table}[!th]
\caption{Significant gradients (2$\sigma$ level) of Theia~216.} 
\label{tab_grad_kpc_summary}
\centering
\begin{tabular}{l|rrrrr}
\hline\hline 
element & Fe & Ca & Cr & Mn & Ni\\
gradient (dex~kpc$^{-1})$ & 0.99 & 2.87 & 0.92 & 1.92 & 0.75\\
\hline
\end{tabular}
\end{table} 

The possibility of strong inhomogeneities at small scales 
is actually supported by the literature.  The interstellar medium (ISM) can be clumpy 
and chemically inhomogeneous down to a size of a few parsecs \citep{decia}, or even
0.1~pc \citep{nasoudi-shoar}. In particular, we cite here one finding of 
\citet{decia}, whose authors measured the metallicity of the ISM 
seen in the line of sight of some OB stars. 
Measurements were taken via two different methods: the
relative method and the F$^*$ method. Both of them are aimed at obtaining a
metallicity corrected for dust depletion\footnote{The relative method derives the overall strength of depletion from the relative abundance of two elements that follow each other in nucleosynthesis but have different refractory properties. On the other hand, the $F^*$ method derives the dust depletion by correlating all the observed abundances and minimizing the residuals with respect to a common factor, $F^*$. All the details can be found in \citet{decia}.}. The stars $\epsilon$ Per 
and $\rho$ Oph A are at a distance of 82 and 139~pc from the Sun, respectively, and
subtend an angle of $\sim 158.8$~deg from each other. This results in a physical
distance of 217~pc between them. The ISMs along their line of sight have metallicities 
that differ by 0.6~dex (for the relative method) and 0.23~dex (for the F$^*$ method),
which corresponds to metallicity gradients of at least 2.7~\dexkpc\ and 1.0~\dexkpc, respectively. 
This leads us to the conclusion that the chemical gradients observed in
Theia~216 are not exceptionally large.
This points in favor of star formation in an inhomogeneous filament.

\section{Discussion}\label{sec_discussion}
Figure~\ref{fig_mean_std} summarizes the chemical features of the 20 objects
under study, while Fig.~\ref{MAD_PM_RV_objects} summarizes their kinematical features.
From these figures we derive the evidence listed below.
\begin{itemize}

\item Fifteen out of 20 objects crowd together in the $-0.10
\leq[Fe/H]\leq0.00$~dex and $0.04\leq\sigma_{[Fe/H]}\leq0.10$~dex range.
They have the same average iron abundance as their test samples ($[Fe/H]\sim
-0.05$~dex) but are distinct for lower abundance dispersions (test
samples have $\sigma_{[Fe/H]}\sim0.19$~dex on average).

\item Most of the objects have a chemical homogeneity comparable to 
OCs (blue points and stars in Fig.~\ref{fig_mean_std} have similar $\sigma_{[Fe_H]}$).

\item The rest of the objects (Theia~908, 823, 1004, 1415,
and 1531) distinguish themselves for having iron abundances
located at the tails of the distribution, with Theia~908 on the metal poor side
($[Fe/H]\sim-0.15$~dex) and the other four objects being the most metal rich
($[Fe/H]>0.05$~dex). It should be noted that all these objects are also the
oldest, with a $log~age$ between 8.62 and 9.4~dex.

\item With an internal velocity dispersion, $MAD_{RV}^{obj}$, in the
1--7~\kmsec\ range, all objects (with the exception of Theia~1415 and Theia~908, 
which have $MAD_{RV}^{obj} \sim 15$ and 9.5~\kmsec, respectively) 
have internal kinematics that are clearly distinct from those of OCs \citep[see][]{madsen, olivares} and the test samples (with $MAD_{RV}^{test}$ being between 12 and 25~\kmsec).

\end{itemize}
These observations depict Galactic strings (but also the three rounder objects
studied by us) that are i) as chemically homogeneous as OCs, ii) kinematically
distinct from (hotter than) OCs, and iii) kinematically colder than the local field. They are
spatially extended and can embed one or more OCs.

 \subsection{The older objects}
While younger objects (but also their field stars) lie in the -0.10--0.00~dex $[Fe/H]$ interval, the five older ones Theia~908, 823, 1004, 1415,
and 1531 do not; they are at the extremes of the metallicity distribution, as mentioned above. This suggests that they may have been born at
different Galactic radii and then had enough time to migrate to the solar
radius. In particular, Theia~908 and Theia~1415
are not associated with any known OC, and their metallicity dispersion is as broad
as the surrounding field stars. While Theia~908 is chemically indistinguishable from field stars, 
Theia~1415 shows chemical inhomogeneity and distinguishes itself for having a 
mean Fe abundance of $[Fe/H]=0.077$~dex, which is much higher than the other strings. 
These features, joined to its relatively old age,
suggest that maybe a moving group migrated from the inner disk,
kinematically triggered  by the Galactic arms, bar, or resonances \citep{quillen}. 
The existence of such migrating objects has been shown by
backward orbit integration of young stellar association by
\citet{quillen2020}. In contrast, the relatively low
[Fe/H] and high [Mg/Fe] of Theia~908 could also be explained by localized infall
of metal-poor gas, a scenario invoked to explain the existence of young,
metal-poor stars in {\it Gaia}-GSP data \citep{spitoni}.
However, given the small statistics of Theia~908 and Theia~1415 (10 stars each) such scenarios have to be corroborated by new data to be collected in the future. For the time being, the possibility that these two objects are merely artifacts cannot be excluded.

The other three objects Theia~1531, Theia~1004, and Theia~823 are the most metal 
rich (all with [Fe/H]$>0.05$~dex) and embed
one OC each (Ruprecht~147, the Hyades, and Mamajek~4, respectively). 
All of these OCs have tidal tails \citep{tarricq, meingast_hyades}; therefore, the object identified by KC19
may be composed of the cluster and its tidal tail. Although Theia~1531 and Theia~1004 are 
classified as roundish objects, while Theia~823 was classified as string by KC19, the tidal disruption process is ongoing for all of them. For such objects (i.e., those with an embedded cluster)
\citet{meingast2021} showed how some OCs appear to be embedded
in extensive coronae co-moving with their central cluster. These can be cluster members that escaped from the OC's gravitational well as part of the evaporation process. Since these stars left the cluster
gravitational well, they are expected to have a velocity dispersion higher
than that of the central cluster. The several objects studied by us exhibiting
chemical homogeneity and internal velocity dispersions on the order of 
1--3 \kmsec\  may be explained with the formation scenarios just outlined above.
Once the evaporation process is completed, the object will appear kinematically hotter than OCs, but not as much as the field stars, chemically homogeneous and extended in space, lacking a central core, matching the features we observe in some of the younger strings studied such as Theia~309 and M1.
A chemically homogeneous object
with no OC association (and no evident core) is a good candidate evaporated 
cluster, or a ``fragment of a cluster,'' as suggested by \citet{gagne}.
On the other hand, the stars' coronae discussed by \citet{meingast2021} can be stars that formed in the neighborhoods of OCs and are still co-moving with them. This scenario is more likely to be found in younger objects because such coronae did not have enough time to disperse in the surrounding star field.

\subsection{The younger objects}

\begin{figure*}[t]
\sidecaption
\includegraphics[width=12cm, clip]{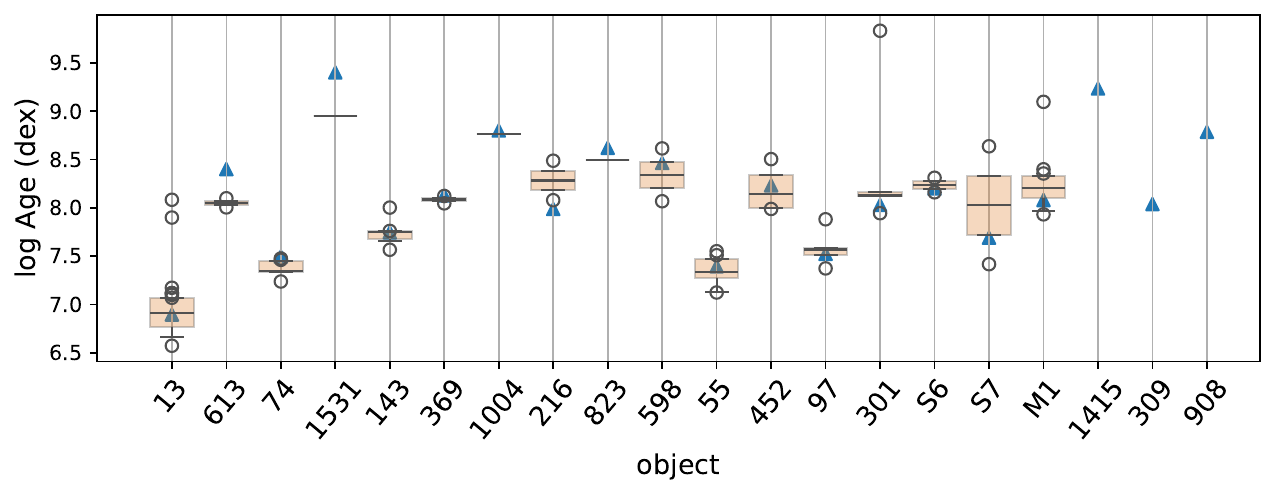}
\caption{Log age of the objects as estimated by KC19 (blue triangles) compared with the ages of the OCs embedded in the objects as given by HR24 (showed as box plot with mean, quartile, and whiskers that span the $3\sigma$ probability level). Open circles are points that lie beyond the $3\sigma$ probability level.} 
\label{ages_OCs_objects}
\end{figure*}

Their relatively young ages (10--400~Myrs) and their chemical homogeneity suggest that they
formed in the same environment as the OCs they embed. This is supported by the embedded OCs ages as estimated by HR24, which are in fair agreement with the object ages as estimated by KC19 (see Fig.~\ref{ages_OCs_objects}).

We highlight the case of Theia~74, which embeds three OCs that have
indistinguishable chemistry (and nearly the same age) and the cases of
Theia~216 and Theia~309, which are spatially close and chemically
indistinguishable. We also note that Theia~143 and Theia~74 
belong to the western part of the stellar structure identified by \citet{wang}, 
while Theia~55 is located in the eastern part of it. Their similar ages ($log~age =$ 7.75, 7.50, and 7.39~dex for Theia~143, Theia~74, and Theia~55, respectively) and their mismatching chemistry are consistent with chemically inhomogeneous complex star formation with different bursts: first Theia 143, then Theia 74, then Theia 55 \citep[similar conclusions were drawn in the Radcliffe wave study by][]{alonso-santiago}.
These three strings also belong to the Vela-Puppis region \citep{cantat-gaudin2019}. These objects seem to have some features in common with hierarchical 
structures such as the LISCA objects \citep{lisca1, lisca2} and the structures in the
Perseus region \citep{perseus}, whose OCs are connected by bridges of stars or
embedded in expanding gas and stellar clouds. 

The case of Theia~216 and its chemical gradients may suggest a formation scenario 
similar to the one proposed by \citet{meingast2021} and \citet{coronado}.
The ISM is known to be organized in filaments \citep{myers, ragan}. A star formation event
occurring in such a filament would generate a group of coeval stars of an elongated
shape. If a chemical gradient were present along the filament,
this would be reflected by the individual chemical composition of the
stars, which would be detected as a chemical gradient along the object, as we
saw for Theia~216. Another gradient formation mechanism can be
sequential star formation along an originally chemically homogeneous filament.
Massive stars formed at one end of the filament would
chemically pollute the nearby segment of the filament and, at the same time, 
trigger star formation events when they die via supernova events, thus generating a gradient 
in chemical abundances and ages along the filament \citep{grosschedl}. On top of these
scenarios, we cannot exclude the possibility that the gradients observed may merely be caused by the presence of two OCs that have different metallicities.
More data are needed to prove the consistence of such formation scenarios. 

\subsection{Dispersal time of these structures}
One peculiarity of many of these elongated structures is that they exhibit
relatively young ages, from a few tens to hundreds of Myrs (see KC19, but also \citealt{meingast}, \citealt{curtis}, and \citealt{tarricq}).
This seems to contradict their
existence. In fact, most of the objects in this work appear to have 
an internal velocity dispersion that would
lead them to dissolve in a time shorter than their age, as pointed out by
\citet{zucker}. In Sect.~\ref{sec_kinematic} we state that \citet{zucker} based their
dispersal time estimate on the internal RV dispersion, which (as we show)
is inflated by velocity outliers misclassified as object members. The way we handled the 
RV dispersions led to lower dispersion values, i.e., to longer dispersal times.  We then calculated the dispersal times of the $v_t$ as $t_{d,v_t} = size/MAD_{v_t}$ (and similarly for RVs $t_{d,RV}$),
where the size is the minimum value among the length, height, and width reported by KC19 (similarly for S6, S7, and M1 reported in R25 and M19). These values are reported in Table~\ref{tab_Age_DispTime}, together with the age/dispersal-time ratio. 

\begin{table}[!th]
\caption{Age (sorted from oldest to youngest), dispersal times $t_{d,v_t}$ and $t_{d,RV}$, and the ratio between age and dispersal times. The latter values are higher than one when the object should not have survived to its evaporation.} 
\label{tab_Age_DispTime}
\centering
\begin{tabular}{l|rrrrr}
\hline\hline 
Theia & age & $t_{d,v_t}$ & $t_{d,RV}$ & age/$t_{d,v_t}$ & age/$t_{d,RV}$ \\
      & (Myr) & (Myr) & (Myr) &  & \\
\hline
1531 & 2517.9 & 3.9 & 4.1 & 642.3 & 609.1\\
1415 & 1716.8 & 4.8 & 1.0 & 359.5 & 1677.6\\
1004 & 631.0 & 5.6 & 2.0 & 112.4 & 317.1\\
908 & 611.1 & 7.2 & 2.1 & 84.6 & 286.6\\
823 & 417.6 & 3.1 & 2.8 & 136.4 & 148.9\\
598 & 290.9 & 6.2 & 2.8 & 46.9 & 103.6\\
613 & 251.4 & 6.6 & 3.3 & 38.3 & 76.0\\
452 & 170.8 & 1.7 & 1.6 & 101.2 & 110.2\\
S6 & 158.5 & 75.6 & 54.1 & 2.1 & 2.9\\
369 & 134.8 & 5.3 & 2.4 & 25.6 & 55.8\\
M1 & 120.2 & 19.8 & 7.0 & 6.1 & 17.2\\
309 & 109.4 & 6.0 & 4.9 & 18.3 & 22.4\\
301 & 107.0 & 6.2 & 4.5 & 17.2 & 23.9\\
216 & 97.3 & 5.9 & 4.8 & 16.5 & 20.2\\
143 & 55.8 & 1.4 & 0.7 & 40.6 & 75.9\\
S7 & 49.0 & 55.2 & 14.4 & 0.9 & 3.4\\
97 & 33.3 & 10.3 & 3.3 & 3.2 & 9.9\\
74 & 31.3 & 6.3 & 3.1 & 5.0 & 10.1\\
55 & 24.8 & 13.3 & 2.9 & 1.9 & 8.7\\
13 & 7.9 & 11.2 & 2.5 & 0.7 & 3.1\\
\hline
\end{tabular}
\end{table} 

Although such ratios are still mostly higher than one (i.e., the structures should not exist)\footnote{Only two strings have ratios, $age/t_{d,v_t}$, smaller than one: Theia~13 and S7.}, their average values ($\sim21$ for $v_t$ and $\sim36$ for RV) are significantly lower than the one indicated by \citet{zucker} ($\sim126$). 
   
To reconcile the discrepancy between age and dispersal time, we propose 
two possible explanations. The first is that
such objects may be much more extended than they appear. If they experienced the 
evaporation process, we expect that the stars that
evaporate first would be found farther from the center than the ones that
evaporate later and that the former may be kinematically heated more than
the latter by encounters with giant molecular clouds or another kinematic
mechanism. At the same time, this expanding corona would become less and less
dense, making it more and more difficult to detect over the field stars.
The detectable part of the structure would be
the central part, which is the last one to lose coherence in the phase space. A larger object would make the dispersal time longer and the ratio $age/t_d$ smaller.\\
The second possible explanation comes from a different interpretation of their existence.
So far, we have discussed the existence of strings by
comparing the dispersing time with their ages. Given that some strings may
be born gravitationally bound as OCs, the dissolving time should not be compared
against the time when they were born (i.e, their age) but against the time at which they become unbound.
The latter would be shorter than the former and in many cases also shorter than the
dispersing time. This argument is fully supported by the works of \citet{miller} and \citet{miret-roig}. The former
studied the Galactic string OCSN-49 \citep[discovered by][]{qin} and found 
a dynamical age of $83\pm1$~Myr (the time the structure becomes unbound) compared to 
an age of 400--600~Myr derived by isochrone fitting, gyrochronology, and lithium content. The latter studied stellar associations younger than $\sim40$~Myrs and found that the ages derived from isochronal and dynamical traceback methods differ by about 5~Myrs. This difference is interpreted as the time needed for the association to blow away the gas it was embedded in and that kept the association gravitationally bound. This finding establishes the difference between the isochronal age (when the stars were born) and the dynamical age (when the stars became gravitationally unbound).

\subsection{Galactic string: A broad definition for a variety of objects}
We establish that the elongated objects we studied can have different origins and evolutions, which are listed below.
\begin{itemize}

\item Theia~908 and Theia~1415 are made of stars whose chemical-abundance distributions are as broad as field stars, making them good candidates to be moving groups formed via dynamical mechanisms \citep{quillen}. 

\item Theia~143, Theia~369, Theia~613, and Theia~598 are objects associated with open clusters that are found to be elongated by their tidal tails \citep{tarricq}. In some cases (e.g., Theia~1004) the tidal tails extend for hundreds of parsecs \citep{meingast_hyades}.  

\item Most of the objects studied embed one or more OCs. We showed that at least in one case (Theia~74) three embedded OCs have similar chemical fingerprints and very similar ages. This structure looks like a star formation hub that can see several OCs born from the same interstellar matter cloud. LISCA objects \citep{lisca1, lisca2} can be the prototype of such structures.

\item Objects that are chemically homogeneous and lack an inner core (Theia~309) may be evaporated clusters \citep{gagne} dissolving in the field after losing the gravity bound.

\item The presence of chemical gradients \citep[Theia~216, but see also Theia~456 by][]{andrews} suggests a series of star formation events that took place along a cold interstellar matter filament \citep{coronado}, where the previous event generated Type II SNe explosions that chemically enriched the nearby interstellar medium and triggered the next star formation event. Such structures would exhibit chemical and age gradients \citep{grosschedl}.

\end{itemize}
As outlined here, there are a variety of ways to build up an elongated stellar structure to stress the need to identify which mechanism generates what event. More precise ages and chemical information will allow us to pursue such an identification.

\section{Summary and conclusions}
In this work, we studied the properties of elongated stellar structures found
in the Galactic disk in recent years and named strings by \citet{kounkel}
\citep[but see also][among others]{meingast,curtis,roser2020, ratzenboeck1}. We selected
20 such objects for which we derived the kinematics from $Gaia$ DR3
astrometry and the chemistry from the GALAH DR4 catalog.
Seventeen of them are classified as strings, while three of them are classified as
compact, not-elongated objects according to \citet{kounkel}.
We studied the mean and standard deviation of their chemical abundances as
well as the internal velocity dispersion of such objects. Chemical abundances 
and internal kinematics were compared with the samples of the local 
field stars surrounding each object (test samples).

Seventeen of these objects embed OCs or MGs \citep[as classified
by][]{hunt-reffert2024} and only three objects have none of them. Fifteen objects
have average metallicity similar to the metallicity of their field stars
(test samples), but they are distinct in terms of the metallicity dispersions,
which are narrow for the objects ($\sigma_{[Fe/H]}<0.1$~dex) and broad for
their surrounding field stars ($\sigma_{[Fe/H]}\sim0.19$~dex; see Fig.~\ref{fig_mean_std}). 
Kinematically, we measured the internal velocity dispersions
(as the median absolute deviation of the tangential velocities $MAD_{v_t}^{OC}$ and the radial velocity $MAD_{RV}^{OC}$) of the embedded OCs or MGs as well as those of the
whole objects ($MAD_{v_t}^{obj}$ and $MAD_{RV}^{obj}$). We found that while the associated OCs/MGs
have all $MAD_{v_t}^{OC}<1$\kmsec\ and $MAD_{RV}^{OC}<3$\kmsec,\, their host objects are
kinematically hotter, with $MAD_{v_t}^{obj}$ and $MAD_{RV}^{obj}$ in the 1--7~\kmsec\ range, where
their surrounding star fields have $MAD_{RV}^{test}$ as high as $\sim26$~\kmsec.

The chemical homogeneity of most of these structures suggests that they
may be remnants of evaporated (or still evaporating) open
clusters or objects that formed in star formation hubs where OCs
form connected by (or embedded in) bridges of stars. Such hubs can be represented by
hierarchical formation structures called LISCA objects \citep{lisca1, lisca2}. 
This seems to be supported by the similarity of the ages of the objects and the OCs embedded in them (see Fig.~\ref{ages_OCs_objects}),
suggesting that the objects and associated OCs may have been born together.
In these cases, 
the resulting strings may preserve differences in chemical abundances 
at different locations of the interstellar cloud 
which could be original or generated by chemical pollution in the neighborhood of massive stars that died 
via Type II SNe \citep{grosschedl}. A possible example of such a formation
scenario is provided by the string Theia~216, which exhibits clues of chemical
gradients for some elements.

We saw that what we generically call strings may actually refer to a variety of structures with different origins and evolutions, from evaporating open clusters to dispersing star formation hubs, and original interstellar matter filaments that experienced star formation along their extension to moving groups kinematically triggered by spiral arms or the Galactic bar. More chemical and
kinematic data are needed to confirm the findings of this work and provide new information to ensure a better definition of such structures. 

The upcoming
$Gaia$ DR4 will provide more data of better quality. Future
surveys such as the WHT Enhanced Area Velocity Explorer (WEAVE,
\citealt{dalton2012,dalton2016,jin2024}) and the 4-metre Multi-Object Spectroscopic
Telescope (4MOST, \citealt{dejong}),
equipped with multi-object spectroscopic instruments, 
will allow us to target such elongated structures and provide precise
chemical abundances and radial velocities for a high number of members. A
more robust statistics will help us to better constrain the physical 
conditions of the locations they originated from.

\section{Data availability}
Table~\ref{tab_abd} is only available in electronic form at the CDS via anonymous
ftp to cdsarc.u-strasbg.fr (130.79.128.5) or via http://cdsweb.u-strasbg.fr/cgi-bin/qcat?J/A+A/.

\begin{acknowledgements}
CB, AV, AB, EC, and DR acknowledge funding from Bando Astrofisica Fondamentale INAF 2023 (Open Clusters and stellar structures in the local Galactic disk, PI A. Vallenari). BC acknowledge funding from ASI grant n. 2025-10-HH.0.
SL acknowledges funding from Bando Astrofisica Fondamentale INAF 2022 (Stellar Clusters in 4MOST, PI Lucatello). AB acknowledges funding from Bando Astrofisica Fondamentale INAF 2022 (High resolution spectroscopy of open clusters, PI Bragaglia). JAS and AF acknowledge funding from the Large Grant INAF 2024 (Spectral Key features of Young stellar objects: Wind-Accretion LinKs Explored in the infraRed, SKYWALKER, PI Frasca). CB thanks Rosanna Sordo for useful discussion.
\end{acknowledgements}

\begin{appendix}

\onecolumn
\section{Table and figures}

\begin{table*}[ht]
\small
\centering
\caption{Chunk of the table that reports means and standard deviations of the chemical abundances for the
studied objects and their test samples for Fe and Si. The complete table
with all the elements is available on CDS.}
\label{tab_abd}

\begin{tabular}{l|rrrrrr|rrrrrr|}

\hline
        & \multicolumn{6}{c|}{[Fe/H]}& \multicolumn{6}{c|}{[Si/H]}\\
Theia N & $mean_{obj}$ & $\sigma_{obj}$ & $N_{obj}$ &  $mean_{test}$ & $\sigma_{test}$ &
$N_{test}$ & $mean_{obj}$ & $\sigma_{obj}$ & $N_{obj}$ &  $mean_{test}$ & $\sigma_{test}$ &
$N_{test}$ \\
\hline
13 & -0.037 & 0.085 & 63 & -0.064 & 0.193 & 2712 & 0.022 & 0.108 & 60 & -0.021 & 0.175 & 2712 \\
613 & -0.068 & 0.064 & 57 & -0.053 & 0.196 & 5490 & 0.003 & 0.074 & 56 & -0.005 & 0.178 & 5490 \\
74 & -0.051 & 0.055 & 55 & -0.047 & 0.188 & 944 & -0.007 & 0.050 & 53 & -0.003 & 0.169 & 944 \\
1531 & 0.061 & 0.050 & 46 & -0.027 & 0.189 & 1266 & 0.084 & 0.039 & 46 & 0.022 & 0.178 & 1266 \\
143 & -0.014 & 0.053 & 38 & -0.011 & 0.198 & 434 & 0.019 & 0.047 & 38 & 0.010 & 0.183 & 434 \\
369 & -0.030 & 0.095 & 34 & -0.060 & 0.191 & 1039 & 0.010 & 0.051 & 34 & 0.005 & 0.171 & 1039 \\
1004 & 0.108 & 0.089 & 26 & -0.064 & 0.159 & 153 & 0.144 & 0.063 & 26 & 0.008 & 0.127 & 153 \\
216 & -0.017 & 0.040 & 26 & -0.022 & 0.183 & 1025 & -0.002 & 0.049 & 26 & 0.023 & 0.167 & 1025 \\
823 & 0.059 & 0.086 & 17 & -0.031 & 0.193 & 1900 & 0.059 & 0.063 & 17 & 0.010 & 0.178 & 1900 \\
598 & -0.082 & 0.098 & 16 & -0.023 & 0.191 & 6398 & -0.054 & 0.100 & 16 & 0.015 & 0.175 & 6398 \\
55 & -0.065 & 0.100 & 12 & -0.056 & 0.192 & 3362 & 0.000 & 0.110 & 11 & -0.019 & 0.175 & 3362 \\
452 & -0.048 & 0.094 & 11 & -0.029 & 0.192 & 12903 & -0.007 & 0.096 & 10 & 0.010 & 0.177 & 12903 \\
97 & -0.079 & 0.073 & 10 & -0.057 & 0.195 & 1999 & -0.004 & 0.081 & 9 & -0.022 & 0.178 & 1999 \\
1415 & 0.077 & 0.195 & 10 & -0.029 & 0.198 & 4576 & 0.111 & 0.212 & 10 & 0.007 & 0.184 & 4576 \\
309 & -0.027 & 0.055 & 10 & -0.03 & 0.188 & 3623 & 0.016 & 0.048 & 10 & 0.015 & 0.172 & 3623 \\
301 & -0.060 & 0.085 & 10 & -0.063 & 0.196 & 14862 & 0.014 & 0.083 & 10 & -0.017 & 0.179 & 14862 \\
908 & -0.148 & 0.252 & 10 & -0.041 & 0.181 & 3481 & -0.050 & 0.182 & 10 & 0.018 & 0.162 & 3481 \\
S6 & -0.060 & 0.077 & 11 & -0.054 & 0.173 & 8112 & 0.021 & 0.080 & 11 & 0.016 & 0.154 & 8112 \\
S7 & -0.101 & 0.059 & 10 & -0.044 & 0.18 & 1543 & 0.033 & 0.117 & 10 & 0.012 & 0.163 & 1543 \\
M1 & -0.087 & 0.076 & 14 & -0.072 & 0.163 & 5582 & 0.026 & 0.054 & 14 & 0.016 & 0.145 & 5582 \\
\hline
\end{tabular}
\end{table*}

\scriptsize{\setlength{\tabcolsep}{1mm}
\begin{table*}[!th]
\centering
\tiny
\caption{Open clusters belonging to objects and features.}
\label{tab:clusters}

\begin{tabular}{rlrrrr|rlrrrr}

\hline\hline 
Theia\tablefootmark{(a)} & OC name & $MAD_{v_t}^{OC}$ & $MAD_{RV}^{OC}$  & $N_{cl}$ & $N_{x}$
& Theia & OC name & $MAD_{v_t}^{OC}$ & $MAD_{RV}^{OC}$  & $N_{cl}$ & $N_{x}$  \\
\hline
13 & ASCC\_18 & 0.53 $\pm$ 0.09 & 1.23 $\pm$ 0.66 & 82 & 64 & 1004 & Melotte\_25 & 0.37 $\pm$ 0.03 & 1.21 $\pm$ 0.1 & 386 & 304\\
13 & ASCC\_19 & 0.24 $\pm$ 0.04 & 1.29 $\pm$ 0.69 & 61 & 54 & 216 & HSC\_2712 & 0.47 $\pm$ 0.16 & 1.29 $\pm$ 0.82 & 29 & 19\\
13 & ASCC\_20 & 0.24 $\pm$ 0.02 & 1.34 $\pm$ 0.57 & 194 & 147 & 216 & UPK\_612 & 0.51 $\pm$ 0.05 & 0.70 $\pm$ 0.20 & 228 & 196\\
13 & ASCC\_21 & 0.23 $\pm$ 0.03 & 2.97 $\pm$ 0.65 & 116 & 98 & 823 & Mamajek\_4 & 1.00 $\pm$ 0.08 & 2.24 $\pm$ 0.23 & 480 & 274\\
13 & Briceno\_1 & 0.23 $\pm$ 0.02 & 2.01 $\pm$ 0.93 & 171 & 147 & 598 & Alessi\_9 & 0.20 $\pm$ 0.03 & 0.84 $\pm$ 0.15 & 110 & 83\\
13 & CWNU\_1054 & 0.31 $\pm$ 0.12 & 4.40 $\pm$ 1.33 & 44 & 22 & 598 & HSC\_2773 & 0.54 $\pm$ 0.16 & 1.44 $\pm$ 3.06 & 28 & 14\\
13 & CWNU\_1072 & 0.28 $\pm$ 0.05 & 1.18 $\pm$ 1.49 & 60 & 38 & 55 & ASCC\_24 & 0.16 $\pm$ 0.07 &     & 18 & 12\\
13 & Collinder\_69 & 0.68 $\pm$ 0.03 & 2.60 $\pm$ 0.59 & 752 & 560 & 55 & CWNU\_1007 & 0.27 $\pm$ 0.04 & 2.87 $\pm$ 0.87 & 105 & 62\\
13 & HSC\_1633 & 0.34 $\pm$ 0.05 & 0.44 $\pm$ 2.07 & 68 & 55 & 55 & CWNU\_1092 & 0.28 $\pm$ 0.10 &     & 27 & 12\\
13 & HSC\_1648 & 0.31 $\pm$ 0.06 & 2.50 $\pm$ 2.23 & 78 & 58 & 55 & CWNU\_1111 & 0.18 $\pm$ 0.03 & 1.48 $\pm$ 0.80 & 40 & 32\\
13 & HSC\_1653 & 0.30 $\pm$ 0.07 &     & 30 & 29 & 55 & HSC\_1403 & 0.25 $\pm$ 0.04 & 2.95 $\pm$ 0.89 & 105 & 77\\
13 & L\_1641S & 0.30 $\pm$ 0.05 &     & 72 & 56 & 55 & HSC\_1481 & 0.20 $\pm$ 0.06 &     & 40 & 16\\
13 & NGC\_1980 & 0.25 $\pm$ 0.02 & 2.84 $\pm$ 0.75 & 364 & 295 & 55 & NGC\_2232 & 0.22 $\pm$ 0.02 & 1.90 $\pm$ 0.57 & 286 & 190\\
13 & NGC\_2068 & 0.77 $\pm$ 0.11 &     & 102 & 77 & 55 & OCSN\_54 & 0.27 $\pm$ 0.06 &     & 36 & 27\\
13 & OC\_0322 & 0.32 $\pm$ 0.09 &     & 53 & 39 & 55 & ZHBJZ\_1 & 0.20 $\pm$ 0.01 & 2.04 $\pm$ 0.56 & 458 & 264\\
13 & OC\_0339 & 0.15 $\pm$ 0.05 &     & 19 & 13 & 452 & Alessi\_24 & 0.41 $\pm$ 0.13 & 2.47 $\pm$ 1.11 & 288 & 33\\
13 & OC\_0356 & 0.22 $\pm$ 0.11 &     & 15 & 11 & 452 & CWNU\_61 & 0.56 $\pm$ 0.19 & 2.68 $\pm$ 1.89 & 53 & 27\\
13 & OCSN\_56 & 0.39 $\pm$ 0.07 & 3.0 $\pm$ 1.81 & 88 & 67 & 452 & CWNU\_513 & 0.34 $\pm$ 0.17 &     & 18 & 6\\
13 & OCSN\_59 & 0.35 $\pm$ 0.08 &     & 60 & 49 & 452 & HSC\_2664 & 0.75 $\pm$ 0.18 &     & 26 & 16\\
13 & OCSN\_61 & 0.20 $\pm$ 0.02 & 2.34 $\pm$ 0.90 & 147 & 94 & 97 & CWNU\_410 & 0.18 $\pm$ 0.05 & 2.59 $\pm$ 1.73 & 59 & 28\\
13 & OCSN\_65 & 0.39 $\pm$ 0.07 & 3.01 $\pm$ 1.73 & 70 & 42 & 97 & CWNU\_1024 & 0.52 $\pm$ 0.05 & 2.33 $\pm$ 0.39 & 336 & 217\\
13 & OCSN\_68 & 0.26 $\pm$ 0.06 & 1.53 $\pm$ 0.95 & 51 & 27 & 97 & LP\_2383 & 0.29 $\pm$ 0.03 & 1.42 $\pm$ 0.41 & 313 & 151\\
13 & OCSN\_70 & 0.22 $\pm$ 0.07 &     & 18 & 12 & 97 & OCSN\_76 & 0.38 $\pm$ 0.09 & 0.56 $\pm$ 1.96 & 119 & 25\\
13 & Sigma\_Orionis & 0.61 $\pm$ 0.07 & 1.59 $\pm$ 2.25 & 181 & 139 & 97 & OCSN\_82 & 0.35 $\pm$ 0.04 & 1.74 $\pm$ 0.71 & 157 & 90\\
13 & UBC\_17a & 0.27 $\pm$ 0.04 & 2.56 $\pm$ 1.38 & 102 & 88 & 301 & CWNU\_536 & 0.25 $\pm$ 0.06 & 0.74 $\pm$ 0.64 & 60 & 41\\
13 & UPK\_402 & 0.24 $\pm$ 0.05 &     & 47 & 29 & 301 & HSC\_1580 & 0.44 $\pm$ 0.25 & 0.85 $\pm$ 0.38 & 66 & 11\\
13 & UPK\_422 & 0.35 $\pm$ 0.05 & 3.24 $\pm$ 0.76 & 239 & 144 & 301 & HSC\_1964 & 0.80 $\pm$ 0.07 & 1.55 $\pm$ 0.42 & 257 & 206\\
613 & HSC\_2336 & 0.38 $\pm$ 0.18 & 1.77 $\pm$ 1.68 & 39 & 18 & 301 & HSC\_1991 & 0.20 $\pm$ 0.09 &     & 17 & 11\\
613 & NGC\_2516 & 0.50 $\pm$ 0.02 & 1.80 $\pm$ 0.11 & 1650 & 1151 & 301 & HSC\_2028 & 0.36 $\pm$ 0.16 &     & 10 & 6\\
74 & Alessi\_36 & 0.23 $\pm$ 0.02 & 1.01 $\pm$ 0.42 & 198 & 161 & S6 & HSC\_2407 & 0.41 $\pm$ 0.07 & 0.95 $\pm$ 0.31 & 81 & 61\\
74 & Collinder\_135 & 0.23 $\pm$ 0.02 & 1.26 $\pm$ 0.26 & 209 & 176 & S6 & HSC\_2705 & 0.77 $\pm$ 0.45 & 0.58 $\pm$ 0.47 & 48 & 6\\
74 & Collinder\_140 & 0.31 $\pm$ 0.03 & 3.45 $\pm$ 0.94 & 203 & 127 & S7 & HSC\_1849 & 0.38 $\pm$ 0.16 & 0.76 $\pm$ 0.39 & 48 & 7\\
74 & NGC\_2547 & 0.34 $\pm$ 0.03 & 1.65 $\pm$ 0.40 & 424 & 303 & S7 & NGC\_2451A & 0.26 $\pm$ 0.02 & 0.97 $\pm$ 0.26 & 307 & 293\\
74 & NGC\_2451B & 0.29 $\pm$ 0.02 & 1.50 $\pm$ 0.30 & 564 & 241 & M1 & CWNU\_1010 & 0.16 $\pm$ 0.05 & 1.82 $\pm$ 0.80 & 65 & 16\\
74 & OC\_0450 & 0.31 $\pm$ 0.02 & 2.18 $\pm$ 0.53 & 398 & 236 & M1 & CWNU\_1018 & 0.26 $\pm$ 0.09 & 0.59 $\pm$ 0.70 & 68 & 21\\
74 & UPK\_535 & 0.26 $\pm$ 0.03 & 1.63 $\pm$ 0.57 & 100 & 69 & M1 & CWNU\_1066 & 0.41 $\pm$ 0.14 & 1.62 $\pm$ 0.88 & 56 & 25\\
1531 & Ruprecht\_147 & 0.31 $\pm$ 0.04 & 0.95 $\pm$ 0.21 & 192 & 146 & M1 & HSC\_396 &     &     & 134 & 1\\
143 & ASCC\_58 & 0.38 $\pm$ 0.05 & 1.68 $\pm$ 0.52 & 236 & 127 & M1 & HSC\_508 & 0.53 $\pm$ 0.17 & 2.68 $\pm$ 1.37 & 65 & 13\\
143 & Alessi\_5 & 0.41 $\pm$ 0.03 & 1.96 $\pm$ 0.35 & 427 & 284 & M1 & HSC\_598 & 0.45 $\pm$ 0.14 & 1.44 $\pm$ 0.75 & 66 & 17\\
143 & BH\_99 & 0.43 $\pm$ 0.03 & 1.88 $\pm$ 0.37 & 483 & 317 & M1 & HSC\_749 & 0.53 $\pm$ 0.11 & 1.64 $\pm$ 0.37 & 146 & 50\\
143 & CWNU\_287 & 0.34 $\pm$ 0.08 & 1.89 $\pm$ 1.19 & 36 & 25 & M1 & HSC\_1448 & 0.60 $\pm$ 0.15 & 0.84 $\pm$ 0.28 & 65 & 23\\
143 & HSC\_2215 & 0.40 $\pm$ 0.24 &     & 148 & 12 & M1 & HSC\_1588 & 0.19 $\pm$ 0.10 &     & 13 & 4\\
143 & Trumpler\_10 & 0.44 $\pm$ 0.02 & 1.51 $\pm$ 0.26 & 614 & 384 & M1 & HSC\_1644 &     &     & 14 & 2\\
369 & HSC\_1580 &     &     & 66 & 2 & M1 & HSC\_1714 & 0.47 $\pm$ 0.16 & 2.01 $\pm$ 1.21 & 58 & 15\\
369 & Melotte\_22 & 0.37 $\pm$ 0.01 & 1.01 $\pm$ 0.09 & 1014 & 758 & M1 & HSC\_1923 & 0.29 $\pm$ 0.20 & 0.66 $\pm$ 0.83 & 69 & 9\\
369 & UPK\_303 & 0.36 $\pm$ 0.15 &     & 261 & 8 & &&&&&\\
\hline

\end{tabular}
\tablefoot{
\tablefoottext{a}{The columns are:
Theia: object name; OC\_name: OC embedded in the object; $MAD_{v_t}^{OC}$: 
tangential velocities internal velocity dispersion of the OC (\kmsec);
$MAD_{RV}^{OC}$: 
internal RV dispersion of the OC (\kmsec); $N_{cl}$: number of
OC members according to HR24; $N_{x}$: number of OC members that are also
member of the host object. }
}
\end{table*}
}

\begin{figure*}[th]
\centering
\includegraphics[width=14cm]{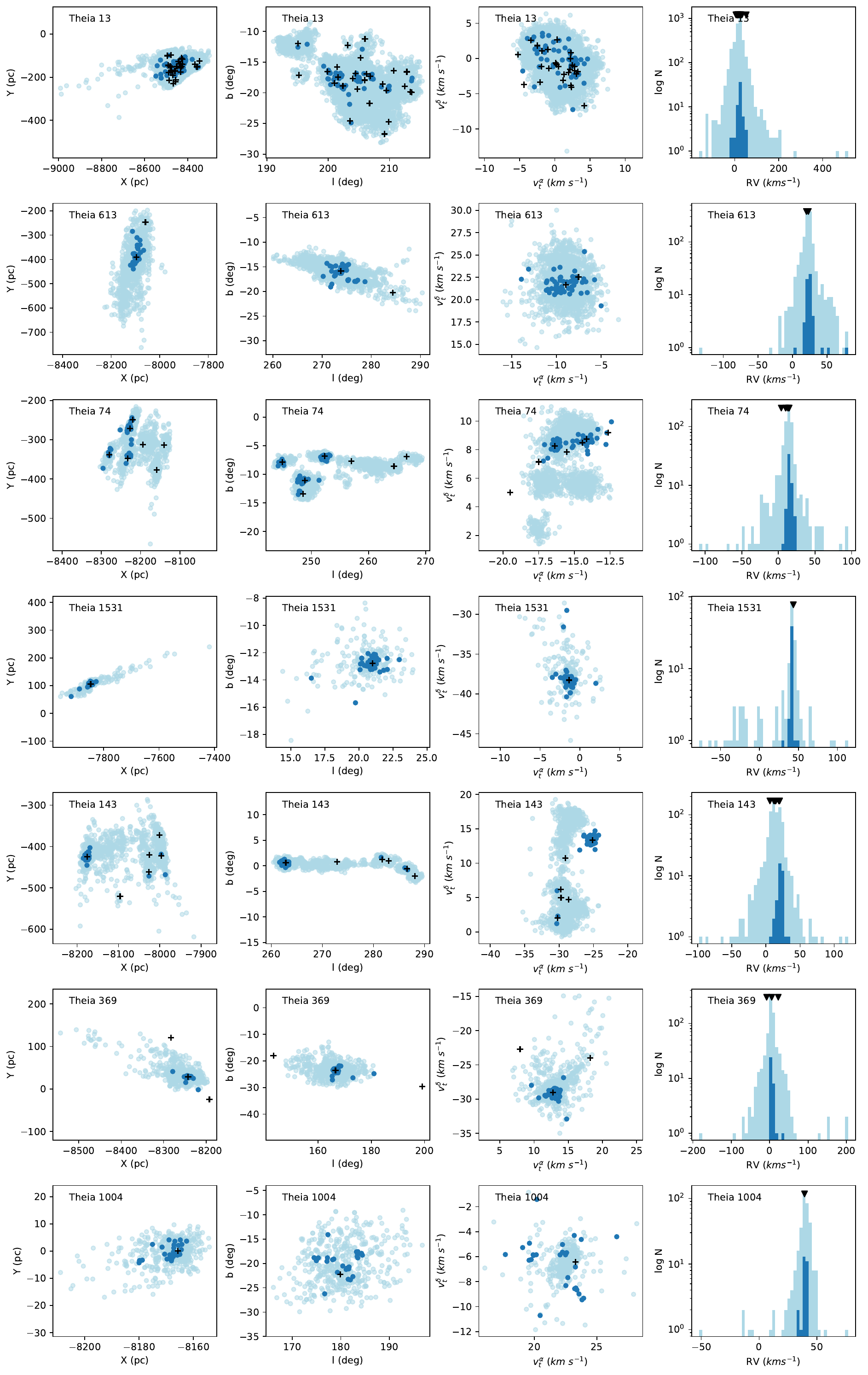}
\caption{Position in Galactocentric and Galactic coordinates, tangential velocities in equatorial
coordinates, distributions in radial velocity for the star members of the objects. The
plus symbols display the positions of the OCs/MGs from
\citet{hunt-reffert2024}. Light blue color refers to the $Gaia$ sample while the dark
blue refers to the GALAH stars subsample of the objects.}
\label{diagnostic1}
\end{figure*}

\begin{figure*}[th]
\centering
\includegraphics[width=14cm]{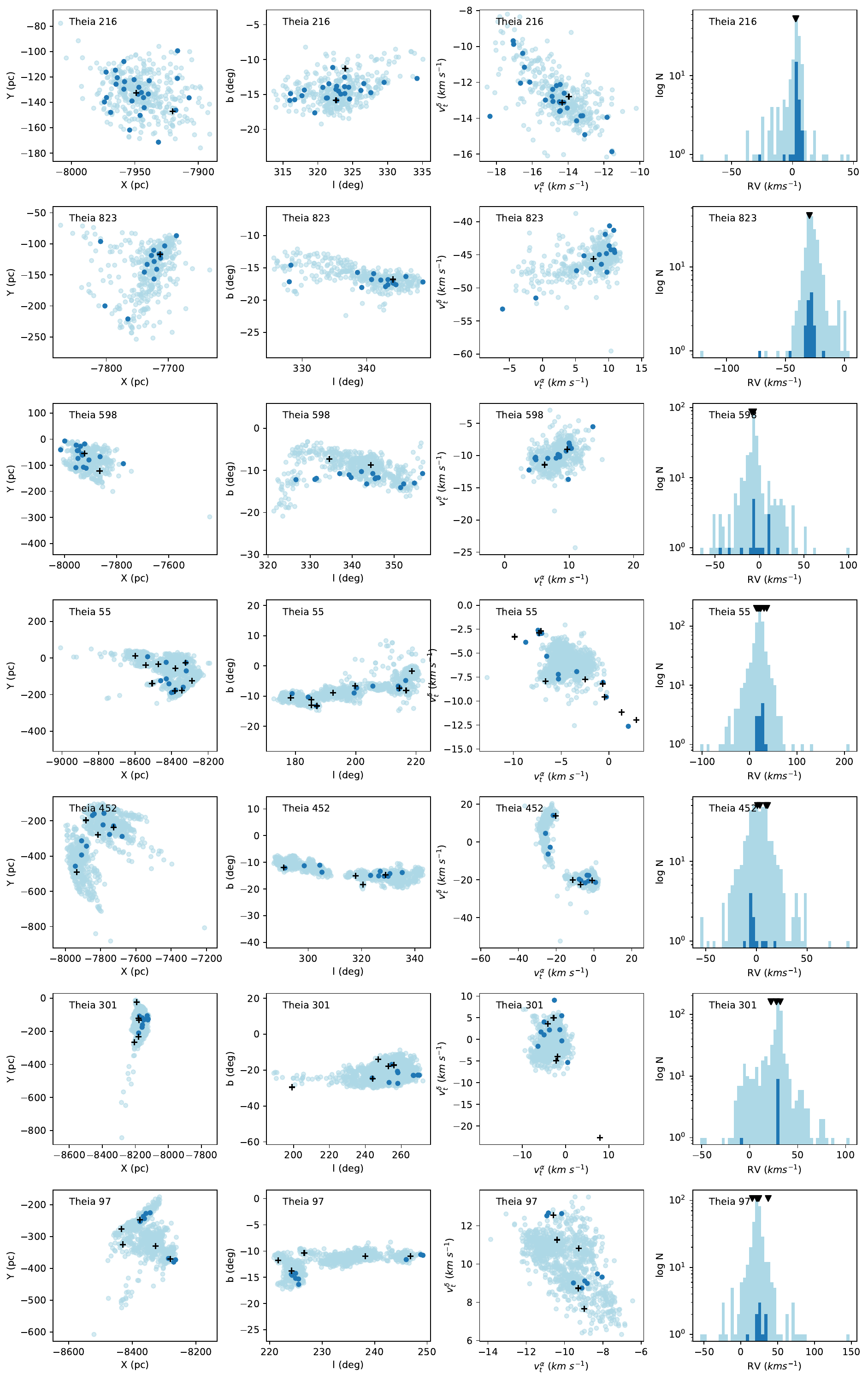}
\caption{As in Fig.~\ref{diagnostic1}.}
\label{diagnostic2}
\end{figure*}

\begin{figure*}[th]
\centering
\includegraphics[width=14cm]{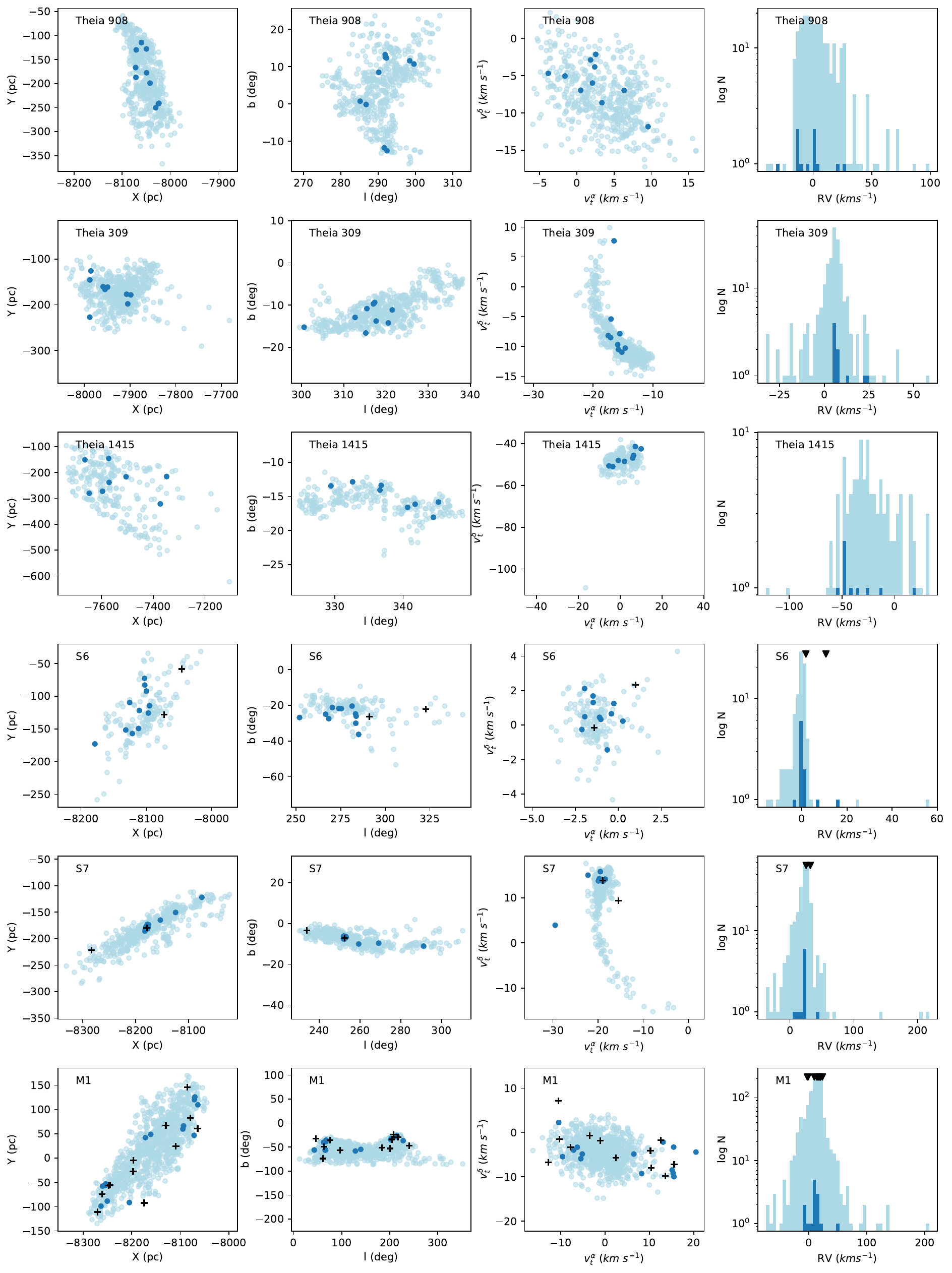}
\caption{As in Fig.~\ref{diagnostic1}.}
\label{diagnostic3}
\end{figure*}

\begin{figure*}[t]
\centering
\includegraphics[width=16cm]{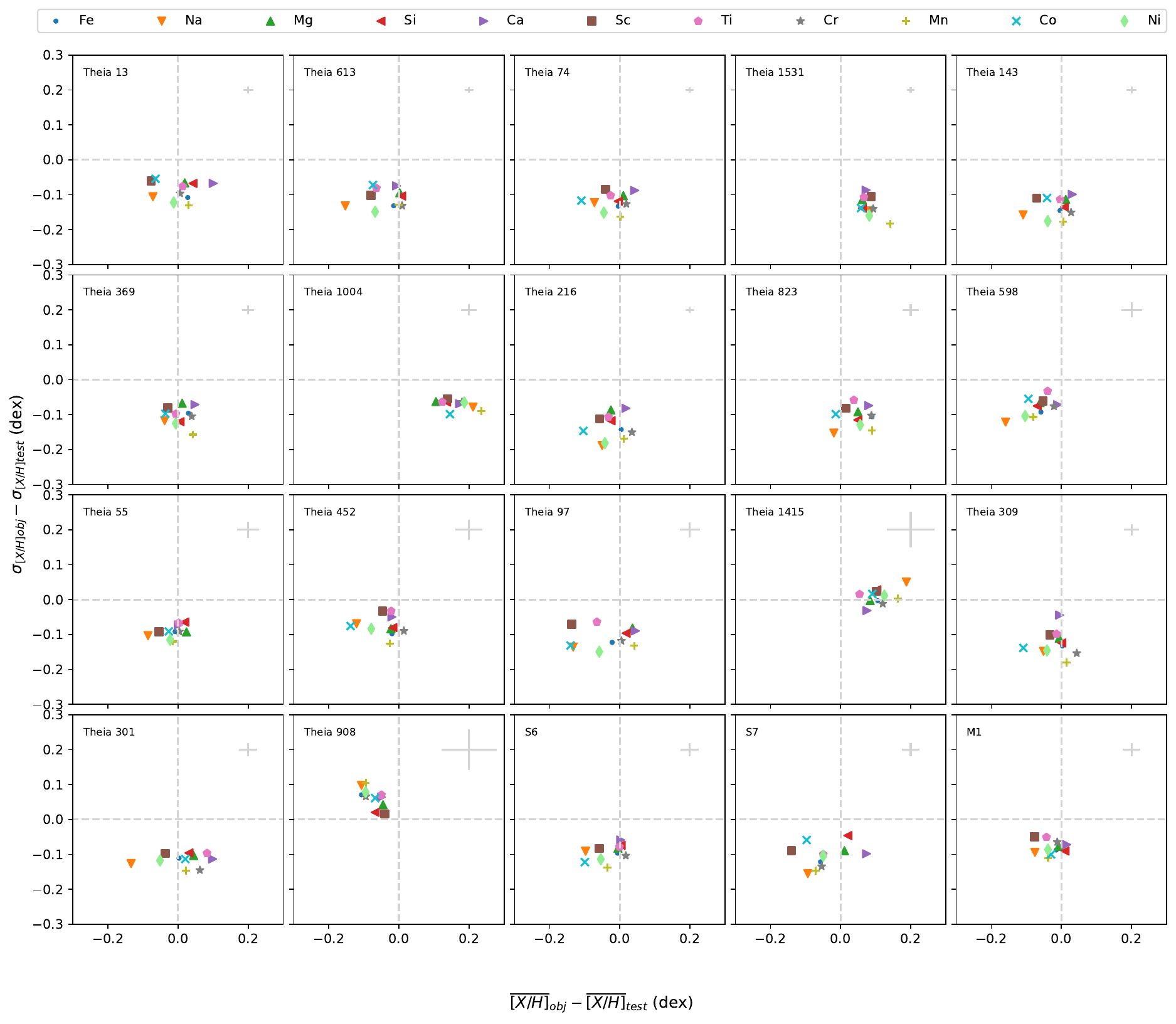}
\caption{Differences of the means (x-axis) and the standard deviation $\sigma$
(y-axis) of the distributions of the elemental abundances between the Theia
object's stars and their test samples. When the values on x-axis are
negative means the Theia objects is metal poorer than its test sample. When
the values on the y-axis are negative means the width of the Theia object's
abundances distributions are smaller than its test sample. Homogeneous
objects have negative values on the y-axis.}
\label{fig_XH_distr_mean_std_diff}
\end{figure*}

\begin{figure*}[t]
\centering
\includegraphics[width=18cm, clip]{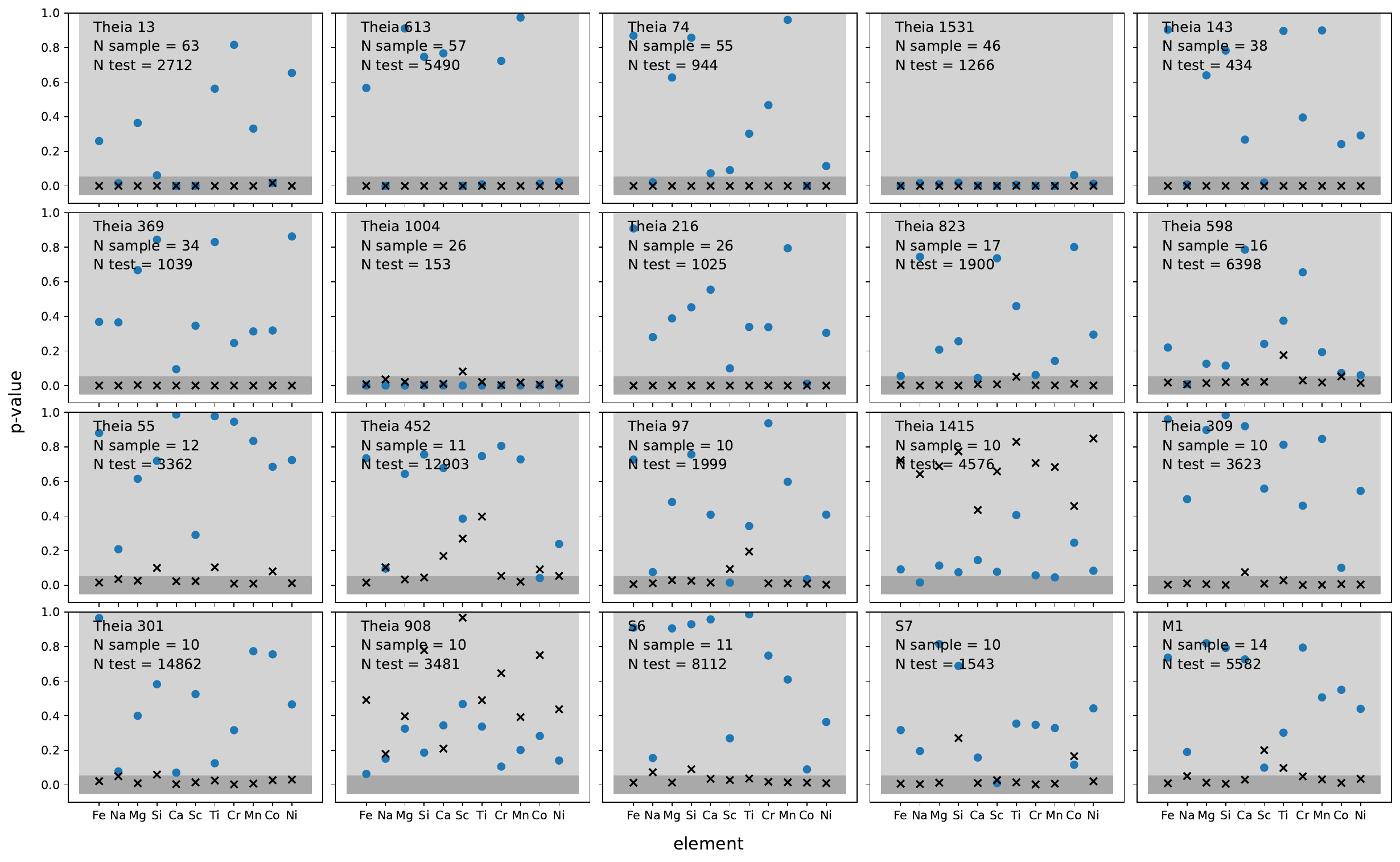}
\caption{t-test and Levene's test p-values (blue points and black crosses, respectively) 
between the object and the test sample abundance
distributions for each element. For two samples belonging to the
same population we expect values larger than 0.05. The region 
for distinct populations is highlighted in dark gray color.}
\label{fig:tt-values}
\end{figure*}

\end{appendix}
\end{document}